\input{psfig}
\input{epsf}
\documentstyle[12pt,frascatiphys]{article}

\makeatletter
\def\@cite#1#2{[{#1\if@tempswa , #2\fi}]}

\renewcommand{\thesection}{\@arabic\c@section.}
\renewcommand{\thesubsection}{\thesection\@arabic\c@subsection}
\makeatother

\def\simge{\mathrel{%
   \rlap{\raise 0.511ex \hbox{$>$}}{\lower 0.511ex \hbox{$\sim$}}}}
\def\simle{\mathrel{
   \rlap{\raise 0.511ex \hbox{$<$}}{\lower 0.511ex \hbox{$\sim$}}}}
 
\def\slashchar#1{\setbox0=\hbox{$#1$}           
   \dimen0=\wd0                                 
   \setbox1=\hbox{/} \dimen1=\wd1               
   \ifdim\dimen0>\dimen1                        
      \rlap{\hbox to \dimen0{\hfil/\hfil}}      
      #1                                        
   \else                                        
      \rlap{\hbox to \dimen1{\hfil$#1$\hfil}}   
      /                                         
   \fi}                                         %
\def\nn{\nonumber}
\def\ts{\thinspace}

\def\ra{\rightarrow}

\def\ol{\bar}

\def\be{\begin{equation}} 
\def\ee{\end{equation}} 
\def\bea{\begin{eqnarray}}
\def\eea{\end{eqnarray}}
\def\ba{\begin{array}}
\def\ea{\end{array}}
\def\chipr{\chi^{\ts \prime}}

\def\CH{{\cal H}}

\def\CO{{\cal O}}

\def\ecm{\sqrt{s}}

\def\atc{\alpha_{TC}}

\def\atro{\alpha_{\tro}}

\def\Ntc{N_{TC}}
\def\suc{SU(3)}

\def\sutc{SU(\Ntc)}

\def\uone{U(1)_1}
\def\utwo{U(1)_2}
\def\uy{U(1)_Y}
\def\suone{SU(3)_1}
\def\sutwo{SU(3)_2}
\def\thw{\theta_W}
\def\kslash{\raise.15ex\hbox{/}\kern-.57em k}
\def\LTC{\Lambda_{TC}}
\def\LETC{\Lambda_{ETC}}
\def\METC{M_{ETC}}
\def\tro{\rho_{T}}
\def\tros{\rho_{T8}^{0}} 
\def\troct{\rho_{T8}} 
\def\tropm{\rho_{T}^\pm}

\def\troz{\rho_{T}^0}
\def\tom{\omega_T}
\def\tpi{\pi_T}
\def\tpipm{\pi_T^\pm}
\def\tpimp{\pi_T^\mp}
\def\tpip{\pi_T^+}
\def\tpim{\pi_T^-}
\def\tpiz{\pi_T^0}
\def\tpipr{\pi_T^{0 \prime}}

\def\octpi{\pi_{T8}}
\def\octpipm{\pi_{T8}^\pm}

\def\octpiz{\pi_{T8}^0}

\def\condtbt{\langle \bar t t\rangle}

\def\condtc{{\langle \ol T T \rangle}_{TC}}
\def\condetc{{\langle \ol T T \rangle}_{ETC}}
\def\tpilq{\pi_{\ol L Q}}

\def\tpind{\pi_{\ol N D}}
\def\tpied{\pi_{\ol E D}}
\def\tpiql{\pi_{\ol Q L}}

\def\jet{{\rm jet}}

\def\mev{{\rm MeV}}
\def\gev{{\rm GeV}}
\def\tev{{\rm TeV}}

\def\ipb{{\rm pb}^{-1}}

\def\ifb{{\rm fb}^{-1}}
\def\half{{\textstyle{ { 1\over { 2 } }}}}

\def\fourth{{\textstyle{ { 1\over { 4 } }}}}

\begin{document}
\title{
\vskip -15mm
\begin{flushright}
\vskip -15mm
{\small BUHEP-00-15\\
hep-ph/0007304\\}
\vskip 5mm
\end{flushright}
{\Large{\bf \hskip 0.38truein TECHNICOLOR 2000}}\\
}
\author{
\centerline{{Kenneth Lane\thanks{lane@physics.bu.edu}}}\\
\centerline{{Department of Physics, Boston University,}}\\
\centerline{{590 Commonwealth Avenue, Boston, MA 02215}}\\
}
\maketitle


\baselineskip=14.5pt
\begin{abstract}
This review is based on lectures on technicolor and extended technicolor
presented at the Frascati Spring School in May 2000. I summarize the
motivation and structure of this theory of dynamical breaking of electroweak
and flavor symmetries. Particular attention is paid to the main
phenomenological obstacles to this picture---flavor--changing neutral
currents, precision electroweak measurements, and the large top--quark
mass---and their proposed resolutions---walking technicolor and
topcolor--assisted technicolor. I then discuss the signatures for technicolor
and the existing and upcoming searches for them at LEP, the Tevatron
Collider, and the Large Hadron Collider. The final section lists some
outstanding theoretical questions.
\end{abstract}
\baselineskip=17pt
\newpage

\section{The Motivation for Technicolor and Extended Technicolor}

The elements of the standard model of elementary particles have been in
place for more than 25~years now. These include the $\suc\otimes SU(2)
\otimes U(1)$ gauge model of strong and electroweak
interactions~\cite{smtheory,smexpt}. And, they include the Higgs mechanism
used to break spontaneously electroweak $SU(2)\otimes U(1)$ down to the
$U(1)$ of electromagnetism~\cite{higgs}. In the standard model, couplings
of the elementary Higgs scalar bosons also break explicitly quark and lepton
chiral--flavor symmetries, giving them hard (Lagrangian) masses. In this
quarter century, the standard model has stood up to the most stringent
experimental tests~\cite{pdg,quigg}. The only indications we have of physics
beyond this framework are the existence of neutrino mixing and, presumably,
masses (though some would say this physics is accommodated within the
standard model); the enormous range of masses, about $10^{12}$, between the
neutrinos and the top quark; the need for a new source of CP--violation to
account for the baryon asymmetry of the universe; the likely presence of cold
dark matter; and, possibly, the cosmological constant.  These hints are
powerful. But they are also obscure, and they do not point unambiguously to
any particular extension of the standard model.

In addition to these experimental facts, considerable theoretical discomfort
and dissatisfaction with the standard model have dogged it from the
beginning. All of it concerns the elementary Higgs boson picture of
electroweak and flavor symmetry breaking---the cornerstone of the standard
model. In particular:

\begin{enumerate}

\item{} Elementary Higgs models provide no dynamical explanation for
electroweak symmetry breaking.

\item{} Elementary Higgs models are unnatural, requiring fine tuning of
parameters to enormous precision.

\item{} Elementary Higgs models with grand unification have a ``hierarchy''
  problem of widely different energy scales.

\item{} Elementary Higgs models are ``trivial''.

\item{} Elementary Higgs models provide no insight to flavor physics.

\end{enumerate}

In nonsupersymmetric Higgs models, there is no explanation why electroweak
symmetry breaking occurs and why it has the energy scale of 1~TeV. The Higgs
doublet self--interaction potential is $V(\phi) = \lambda\ts (\phi^\dagger
\phi - v^2)^2$, where $v$ is the vacuum expectation of the Higgs field $\phi$
{\it when} $v^2 \ge 0$. Its experimental value is $v = 2^{-1/4} G_F^{-1/2} =
246\,\gev$. But what dynamics makes $v^2 > 0$? What dynamics sets its
magnitude?  In supersymmetric Higgs models, the large top--quark Yukawa
coupling drives $v^2$ positive, but this just replaces one problem with
another or, to be generous, replaces two with one.

Elementary Higgs boson models are unnatural. The Higgs boson's mass, $M^2_H =
2 \lambda v^2$ is {\it quadratically} unstable against radiative
corrections~\cite{natural}. Thus, there is no natural reason why $M_H$ and
$v$ should be much less than the energy scale at which the essential physics
of the model changes, e.g., a unification scale or the Planck scale of
$10^{16}\,\tev$. To make $M_H$ very much less that $M_P$, say 1~TeV, the
bare Higgs mass must be balanced against its radiative corrections to the
fantastic precision of a part in $M^2_P/M^2_H \sim 10^{32}$.

In grand--unified Higgs boson models, supersymmetric or not, there are two
very different scales of gauge symmetry breaking, the GUT scale of about
$10^{16}\,\gev$ and the electroweak scale of a few 100~GeV. This hierarchy is
put in by hand, and must be maintained by unnaturally--fine tuning in ordinary
Higgs models, or by the ``set it and forget it'' feature of supersymmetry.

Taken at face value, elementary Higgs boson models are free field
theories~\cite{trivial}. To a good approximation, the self--coupling
$\lambda(\mu)$ of the minimal one--doublet Higgs boson at an energy scale
$\mu$ is given by
\be\label{eq:lamtriv}
\lambda(\mu) \cong {\lambda(\Lambda) \over {1 + (24 /16 \pi^2)\ts
\lambda(\Lambda) \ts \log (\Lambda /\mu)}} \ts.
\ee
This coupling vanishes for all $\mu$ as the cutoff $\Lambda$ is taken to
infinity, hence the description ``trivial''. This feature persists in a
general class of two--Higgs doublet models~\cite{rscdk} and it is probably
true of all Higgs models. Triviality really means that elementary--Higgs
Lagrangians are meaningful only for scales $\mu$ below some cutoff
$\Lambda_\infty$ at which new physics sets in. The larger the Higgs couplings
are, the lower the scale $\Lambda_\infty$. This relationship translates into
the so--called triviality bounds on Higgs masses. For the minimal model, the
connection between $M_H$ and $\Lambda_\infty$ is
\be\label{eq:Mtriv}
M_H(\Lambda_\infty) \cong \sqrt{2 \lambda(M_H)} \ts v = {2 \pi v \over
{\sqrt{3 \log (\Lambda_\infty/M_H)}}} \ts.
\ee
Clearly, the cutoff has to be greater than the Higgs mass for the effective
theory to have some range of validity. From lattice--based
arguments~\cite{trivial}, $\Lambda_\infty \simge 2 \pi M_H$. Since $v$ is
fixed at 246~GeV in the minimal model, this implies the triviality bound $M_H
\simle 700\,\gev$.~\footnote{Precision electroweak measurements suggesting
that $M_H < 200\,\gev$ do not take into account additional interactions that
occur if the Higgs is heavy and the scale $\Lambda$ relatively low. Chivukula
and Evans have argued that these interactions allow $M_H = 400$--$500\,\gev$
to be consistent with the precision measurements~\cite{rscne}.} If the
standard Higgs boson were to be found with a mass this large or larger, we
would know for sure that additional new physics is lurking in the range of a
few~TeV. If the Higgs boson is light, less than 200--300~GeV, as it is
expected to be in supersymmetric models, this transition to a more
fundamental theory may be postponed until very high energy, but what lies up
there worries us nonetheless.

Finally, in all elementary Higgs models, supersymmetric or not, every aspect
of flavor is completely mysterious, from the primordial symmetry defining the
number of quark and lepton generations to the bewildering patterns of flavor
breaking. The presence of Higgs bosons has no connection to the existence of
multiple identical fermion generations. The flavor--symmetry breaking Yukawa
couplings of Higgs bosons to fermions are arbitrary free parameters, put in
by hand. As far as we know, it is a logically consistent state of affairs
that we may not understand flavor until we understand the physics of the
Planck scale. I do not believe this. And, I cannot see how this problem, more
pressing and immmediate than any other save electroweak symmetry break
itself, can be so cavalierly set aside by those pursuing the ``theory of
everything''.~\footnote{This is not quite fair. In the early days of the
second string revolution, in the mid 1980s, there was a great deal of hope
and even expectation that string theory would provide the spectrum---quantum
numbers and masses---of the quarks and leptons. Those string pioneers and
their descendants have learned how hard the flavor problem is.}

The dynamical approach to electroweak and flavor symmetry breaking known as
technicolor (TC)~\cite{tc,kltasi,rscreview} and extended technicolor
(ETC),~\cite{etceekl,etcsd} emerged in the late 1970s in response to these
shortcomings of the standard model. This picture was motivated first of all
by the premise that {\it every} fundamental energy scale should have a
dynamical origin. Thus, the weak scale embodied in the Higgs vacuum
expectation value $v = 246\,\gev$ should reflect the characteristic energy of
a new strong interaction---technicolor---just as the pion decay constant
$f_\pi = 93\,\mev$ reflects QCD's scale $\Lambda_{QCD} \sim 200\,\mev$. For
this reason, I write $F_\pi = 2^{-1/4} G_F^{-1/2} = 246\,\gev$ to emphasize
that this quantity has a dynamical origin.

Technicolor, a gauge theory of fermions with no elementary scalars, is
modeled on the precedent of QCD: The electroweak assignments of quarks to
left--handed doublets and right--handed singlets prevent their bare mass
terms. Thus, if there are no elementary Higgses to couple to, quarks have a
large chiral symmetry, $SU(6)_L \otimes SU(6)_R$ for three generations. This
symmetry is spontaneously broken to the diagonal (vectorial) $SU(6)$ subgroup
when the QCD gauge coupling grows strong near $\Lambda_{QCD}$. This produces
35 massless Goldstone bosons, the ``pions''. According to the Higgs
mechanism---whose operation requires no {\it elementary} scalar
bosons~\cite{jjcn}---this yields weak boson masses of $M_W = M_Z\cos\thw =
\half \sqrt{3} g f_\pi \simeq 50\,\mev$~\cite{tc}. These masses are 1600 times
too small, but they do have the right ratio. Suppose, then, that there are
technifermions belonging to a complex representation of a technicolor gauge
group (taken to be $\sutc$) whose coupling $\atc$ becomes strong at $\LTC =
100$s of GeV. If, like quarks, technifermions form left--handed doublets and
right--handed singlets under $SU(2) \otimes U(1)$, then they have no bare
masses. When $\atc$ becomes strong, the technifermions' chiral symmetry is
spontaneously broken, Goldstone bosons appear, three of them become the
longitudinal components of $W^\pm$ and $Z^0$, and the masses become $M_W =
M_Z\cos\thw = \half g F_\pi$. Here, $F_\pi \sim \LTC$ is the decay constant
of the linear combination of the absorbed ``technipions''. Thus, technicolor
provides a dynamical basis for electroweak symmetry breaking, one that is
based on the familiar and well--understood precedent of QCD.

Technicolor, like QCD, is asymptotically free. This solves in one stroke the
naturalness, hierarchy, and triviality problems. The mass of all
ground--state technihadrons, including Higgs--like scalars (though that
language is neither accurate nor useful in technicolor) is of order $\LTC$ or
less. There are no large renormalizations of bound state masses, hence no
fine--tuning of parameters. If the technicolor gauge symmetry is embedded at
a very high energy $\Lambda$ in some grand unified gauge group with a
relatively weak coupling, then the characteristic energy scale $\LTC$---where
the coupling $\atc$ becomes strong enough to trigger chiral symmetry
breaking---is naturally exponentially smaller than $\Lambda$. Finally,
asymptotically free field theories are nontrivial. A minus sign in the
denominator of the analog of Eq.~(\ref{eq:lamtriv}) for $\atc(\mu)$ prevents
one from concluding that it tends to zero for all $\mu$ as the cutoff is
taken to infinity. No other scenario for the physics of the TeV scale solves
these problems so neatly. Period.

Technicolor alone does not address the flavor problem. It does not tell us
why there are multiple generations and it does not provide explicit breaking
of quark and lepton chiral symmetries. Something must play the role of Higgs
bosons to communicate electroweak symmetry breaking to quarks and leptons.
Furthermore, in all but the minimal TC model with just one doublet of
technifermions, there are Goldstone bosons, technipions $\tpi$, in addition
to $W_L^\pm$ and $Z_L^0$. These must be given mass and their masses must be
more than 50--100~GeV for them to have escaped detection. Extended
technicolor (ETC) was invented to address {\it all} these aspects of flavor
physics~\cite{etceekl}. It was also motivated by the desire to make flavor
understandable at energies well below the GUT scale solely in terms of {\it
gauge} dynamics of the kind that worked so neatly for electroweak symmetry
breaking, namely, technicolor.  Let me repeat: the ETC approach is based on
the gauge dynamics of fermions only. There can be no elementary scalar fields
to lead us into the difficulties technicolor itself was invented to escape.

\section{Dynamical Basics}

In extended technicolor, ordinary $SU(3)$ color, $\sutc$ technicolor, and
flavor symmetries are unified into the ETC gauge group, $G_{ETC}$.
Thus, we understand flavor, color, and technicolor as subsets of the quantum
numbers of extended technicolor. Technicolor and color are exact gauge
symmetries. Flavor gauge symmetries are broken at a one or more high energy
scales $\LETC \simeq \METC/g_{ETC}$ where $\METC$ is a typical flavor gauge
boson mass.

In these lectures, I assume that $G_{ETC}$ commutes with
electroweak $SU(2)$. In this case, it must {\it not} commute with electroweak
$U(1)$, i.e., some part of that $U(1)$ must be contained in
$G_{ETC}$. Otherwise, there will be very light pseudoGoldstone bosons which
behave like classical axions and are ruled out
experimentally~\cite{etceekl,CPreview}. More generally, all
fermions---technifermions, quarks, and leptons---must form no more than four
{\it irreducible} ETC representations: two equivalent ones for left--handed
up and down--type fermions and two inequivalent ones for right--handed up and
down fermions (so that up and down mass matrices are not identical). In other
words, ETC interactions explicitly break all {\it global} flavor symmetries
so that there are no very light pseudoGoldstone bosons or
fermions.~\footnote{I leave neutrinos out of this discussion. Their very
light masses are not yet understood in the ETC framework.}

The energy scale of ETC gauge symmetry breaking is high, well above the TC
scale of 0.1--1.0~TeV, into $SU(3) \otimes \sutc$. The broken gauge
interactions, mediated by massive ETC boson exchange, give mass to quarks and
leptons by connecting them to technifermions (Fig.~1a). They give mass to
technipions by connecting technifermions to each other (Fig.~1b).

The graphs in Figs.~1 are convergent: The changes in chirality imply
insertions on the technifermion lines of the momentum--dependent dynamical
mass, $\Sigma(p)$. This function falls off as $1/p^2 \ts (\log(p/\LTC))^c$ in
an asymptotically free theory at weak coupling and, in any case, at least as
fast as $1/p$~\cite{kdlhdp,agchmg}. For such a power law, the dominant
momentum running around the loop is $\METC$. Then, the operator product
expansion tells us that the generic quark or lepton mass and technipion mass
are given by the expressions
\begin{figure}[t]
 \vspace{6.0cm}
\includegraphics{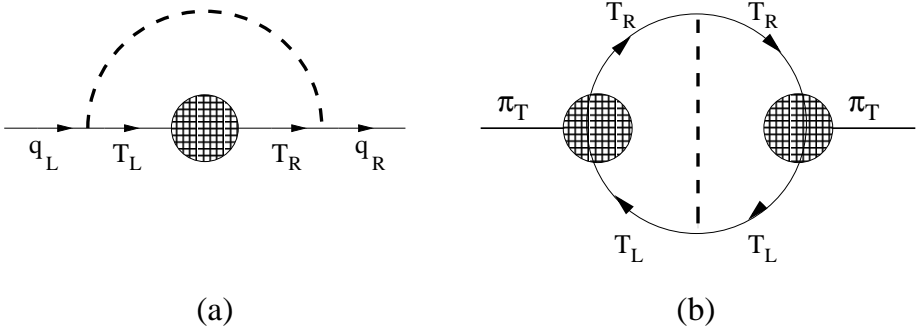}
\vskip-1.5truecm
 \caption{\it
   Graphs for ETC generation of masses for (a) quarks and leptons and (b)
   technipions. The dashed line is a massive ETC gauge boson. Higher--order
   technicolor gluon exchanges are not indicated; from Ref.~\cite{etceekl}.
    \label{fig1} }
\end{figure}
\bea\label{eq:masses}
& & m_q(\METC) \simeq m_\ell(\METC)  \simeq {g_{ETC}^2 \over
{M_{ETC}^2}} \langle \ol T_L T_R \rangle_{ETC} \ts; \\
& & F^2_T M^2_{\pi_T}  \simeq 2\ts {g^2_{ETC} \over {M^2_{ETC}}}
\ts \langle \ol T_L T_R \ol T_R T_L \rangle_{ETC} \ts.
\eea
Here, $m_q(\METC)$ is the quark mass renormalized at $\METC$. It is a hard
mass in that it scales like one for energies below $\METC$. Above that, it
falls off more rapidly, like $\Sigma(p)$. The technipion decay constant
$F_T = F_\pi/\sqrt{N}$ in TC models containing $N$ identical electroweak
doublets of color--singlet technifermions. The vacuum expectation values
$\langle \ol T_L T_R \rangle_{ETC}$ and $\langle \ol T_L T_R \ol T_R T_L
\rangle_{ETC}$ are the bilinear and quadrilinear technifermion condensates
renormalized at $\METC$. The bilinear condensate is related to the one
renormalized at $\LTC$, expected by scaling from QCD to be
\be\label{eq:ctc}
\langle \ol T_L T_R \rangle_{TC} = \half \condtc \simeq 2 \pi F^3_T \ts,
\ee
by the equation
\be\label{eq:condrenorm}
\condetc = \condtc \ts \exp\left(\int_{\LTC}^{M_{ETC}} \ts {d \mu
\over {\mu}} \ts \gamma_m(\mu) \right) \ts.
\ee
The anomalous dimension $\gamma_m$ of the operator $\ol T T$ is given in
perturbation theory by
\be\label{eq:gmm}
\gamma_m(\mu) = {3 C_2(R) \over {2 \pi}} \atc(\mu) + O(\atc^2) \ts,
\ee
where $C_2(R)$ is the quadratic Casimir of the technifermion
$\sutc$--representation $R$. For the fundamental representation of $\sutc$,
$C_2(\Ntc) = (\Ntc^2-1)/2\Ntc$. Finally, in the large--$\Ntc$ approximation
(which will be questionable in the walking technicolor theories we discuss
later, but which we adopt anyway for rough estimates)
\be\label{eq:large_n}
\langle \ol T_L T_R \ol T_R T_L \rangle_{ETC} \simeq \langle \ol T_L
T_R\rangle_{ETC} \ts  \langle \ol T_R T_L \rangle_{ETC} = \fourth \condetc^2
\ts.
\ee

We can obtain an estimate of $\METC$ if we assume that technicolor is
QCD--like. In that case, its asymptotic freedom sets in quickly (or
``precociously'') at energies above $\LTC$ and $\gamma_m(\mu) \ll 1$ for
$\mu$ greater than a few times $\LTC$. Then Eq.~(\ref{eq:ctc}) applies to
$\condetc$. For $N$ technidoublets, the ETC scale required to generate
$m_q(\METC) \simeq 1\,\gev$ is
\be\label{eq:letc}
\Lambda_{ETC} \equiv {M_{ETC} \over {g_{ETC}}} \simeq \sqrt{{2 \pi F_\pi^3
\over {m_q N^{3/2}}}} \simeq {10\,\tev\over{N^{3/4}}} \ts.
\ee
This is pretty low, but the estimate is rough. The typical technipion mass
implied by this ETC scale is
\be\label{eq:mpi}
M_{\tpi} \simeq {\condtc \over{\sqrt{2}\Lambda_{ETC} F_T}} \simeq
{55\,\gev\over{N^{1/4}}} \ts.
\ee

Finally, some phenomenological basics: In any model of technicolor, one
expects bound technihadrons with a spectrum of mesons paralleling what we see
in QCD. The principal targets of collider experiments are the spin--zero
technipions and spin--one isovector technirhos and isoscalar techniomegas. In
the minimal one--technidoublet model ($T = (T_U,T_D)$), the three technipions
are the longitudinal compononents $W_L$ of the massive weak gauge bosons.
Susskind~\cite{tc} pointed out that the analog of the QCD decay $\rho \ra
\pi\pi$ is $\tro \ra W_L W_L$. In the limit that $M_{\tro} \gg M_{W,Z}$, the
equivalence theorem states that the amplitude for $\tro \ra W_L W_L$ has the
same form as the one for $\rho \ra \pi\pi$. If we scale technicolor from QCD
and use large--$\Ntc$ arguments, it is easy to estimate the strength of this
amplitude and the $\tro$ mass and decay rate~\cite{ehlq}:
\bea\label{eq:minrhot}
&&M_{\tro} = \sqrt{{3\over{\Ntc}}}\ts {F_\pi\over{f_\pi}} M_\rho
\simeq 2 \sqrt{{3\over{\Ntc}}}\,\tev \ts, \nn\\
&&\Gamma(\tro \ra W_L W_L ) = {2\atro p_W^3\over{3 M^2_{\tro}}} \simeq 500
\left({3\over{N_{TC}}} \right)^{3/2}\,\gev  \ts.
\eea
Here, the naive scaling argument gives $\atro = (3/\Ntc) \alpha_\rho$ where
$\alpha_\rho = 2.91$.

In the minimal model, a very high energy collider, such as the ill--fated
Superconducting Super Collider (SSC) or a $2\,\tev$ linear collider, is
needed to discover the lightest technihadrons.\footnote{It is possible that,
like the attention paid to discovering the minimal standard model Higgs
boson, this emphasis on the $W_L W_L$ decay mode of the $\tro$ is somewhat
misguided~\cite{kltasi}.  Since the minimal $\tro$ is so much heavier than
$2M_W$, this mode may be suppressed by the high $W$--momentum in its decay
form factor. Then, $\tro$ decays to four or more weak bosons may be
competitive or even dominate. This means that the minimal $\tro$ may be wider
than indicated in Eq.~(\ref{eq:minrhot}) and, in any case, that its decays
are much more complicated than previously thought. Furthermore, walking
technicolor~\cite{wtc}, discussed below, implies that the spectrum of
technihadrons cannot be exactly QCD--like. Rather, there must be something
like a tower of technirhos extending almost up to $\METC \simge$ several
100~TeV. Whether or not these would appear as discernible resonances is an
open question~\cite{hemc}. All these remarks apply as well to the isoscalar
$\tom$ and its excitations.} In nonminimal models, where $N \ge 2$, the
signatures of technicolor ought to be accessible at the Large Hadron Collider
(LHC) and at a comparable lepton collider. We shall argue later that
technicolor signatures are even likely to be within reach of the Tevatron
Collider in Run~II!~\footnote{Run~II of the Tevatron Collider begins in
Spring 2001. The first stage, Run~IIa, is intended to collect $2\,\ifb$ of
data with significantly enhanced CDF and D\O\ detectors featuring new silicon
tracking systems. It is planned that, after a brief shutdown to replace
damaged silicon, Run~IIb will bring the total data sets for each detector to
$15\,\ifb$ or more before the LHC is in full swing in 2006 or so.} Before we
can do that, however, we must face the obstacles to technicolor dynamics and
see how they are overcome.

\section{Dynamical Perils}

Technicolor and extended technicolor are challenged by a number of
phenomenological hurdles, but the most widely cited causes of the ``death of
technicolor'' are flavor--changing neutral current interactions
(FCNC)~\cite{etceekl,ellisfcnc}, precision measurements of electroweak
quantities (STU)~\cite{pettests}, and the large mass of the top quark. We
discuss these in turn.~\footnote{Much of the discussion here on FCNC and STU
is a slightly updated version of that appearing in my 1993 TASI
lectures~\cite{kltasi}.}

\subsection{\it {\underbar{Flavor--Changing Neutral Currents}}}

Extended technicolor interactions are expected to have flavor--changing
neutral currents involving quarks and leptons. The reason for this is simple:
Realistic quark mass matrices require ETC transitions between different
flavors: $q \ra T \ra q'$. Thus, there must be ETC currents of the form $\ol
q'_{L,R} \ts \gamma_\mu \ts T_{L,R}$ and $\ol T_{L,R} \ts \gamma_\mu \ts
q_{L,R}$; their commutator algebra includes the ETC currents $\ol q'_{L,R}
\ts \gamma_\mu \ts q_{L,R}$, and ETC interactions necessarily produce $\ol q
q \ol q q$ operators at low energy. Similarly, there will be $\ol q q \ol
\ell \ell$ and $\ol \ell \ell \ol \ell \ell$ operators. Even if these
interactions are electroweak--eigenstate conserving (or
generation--conserving), they will induce FCNC four--fermion operators after
diagonalization of mass matrices and transformation to the mass--eigenstate
basis. No satisfactory GIM mechanism has ever been found that eliminates
these FCNC interactions~\cite{etcgim}.

The most stringent constraint on ETC comes from $\vert \Delta S \vert = 2$
interactions. Such an interaction has the generic form
\be\label{eq:dstwo}
\CH'_{\vert \Delta S \vert = 2} = {g^2_{ETC} \ts V^2_{ds} \over
{M^2_{ETC}}} \ts\ts \ol d \ts \Gamma^\mu s \ts\ts \ol d \ts \Gamma'_\mu s +
{\rm h.c.}
\ee
Here, $V_{ds}$ is a mixing--angle factor; it may be complex and seems
unlikely to be much smaller in magnitude than the Cabibbo angle, say $0.1
\simle \vert V_{ds} \vert \simle 1$. The matrices $\Gamma_\mu$ and
$\Gamma'_{\mu}$ are left-- and/or right--chirality Dirac matrices. I shall
put $\Gamma_\mu, \ts \Gamma'_\mu = \half \gamma_\mu \ts (1 - \gamma_5)$ and
count the interaction twice to allow for different chirality terms in
$\CH'_{\vert \Delta S \vert = 2}$. The contribution of this interaction to
the $K_L - K_S$ mass difference is then estimated to be
\bea\label{eq:klks}
(\Delta M_K)_{ETC} &\equiv& 2{\rm Re}(M_{12})_{ETC} 
= {4g^2_{ETC} \ts {\rm Re}(V^2_{ds}) \over 8M_{K} {M^2_{ETC}}} \ts \langle
K^0 \vert \ol d \ts \gamma^\mu (1 - \gamma_5) s \ts \ol d \ts \gamma_\mu (1 -
\gamma_5) s \vert \ol K^0 \rangle \nn\\
&\simeq& {g^2_{ETC} \ts {\rm Re}(V^2_{ds}) \over {M^2_{ETC}}} \ts
f^2_K M_K \ts,
\eea
where I used the vacuum insertion approximation with $\langle \Omega
\vert \ol d \gamma_\mu \gamma_5 s\vert \ol K^0(p) \rangle = i \sqrt{2} f_K
p_\mu$ with $f_K \simeq 110\,\mev$. This ETC contribution must be less than
the measured mass difference, $\Delta M_K = 3.5 \times 10^{-18}\,\tev$. This
gives the limit
\be\label{eq:dmlimit}
{M_{ETC} \over {g_{ETC} \ts \sqrt{{\rm Re}(V^2_{ds})}}} \simge 1300\,\tev
\ts.
\ee
If $V_{ds}$ is complex, $\CH'_{\vert \Delta S \vert = 2}$ contributes to the
imaginary part of the $K^0 - \ol K^0$ mass matrix. Using ${\rm Im}(M_{12}) =
\sqrt{2} \Delta M_K |\epsilon| \simeq 1.15\times 10^{-20}\,\tev$, the limit
is
\be\label{eq:epslimit}
{M_{ETC} \over {g_{ETC} \ts \sqrt{{\rm Im}(V^2_{ds})}}} \simge 16000\,\tev
\ts.
\ee

If we use these large ETC masses and scale the technifermion condensates in
Eqs.~(3,4) from QCD, i.e., assume the anomalous dimension $\gamma_m$ is small
so that $\langle \ol T T \ol T T\rangle_{ETC} \simeq \condetc^2 \simeq
\condtc^2 \simeq (4 \pi F^3_T)^2$, we obtain quark and lepton and technipion
masses that are 10--1000 times too small, depending on the size of
$V_{ds}$. This is the FCNC problem. It is remedied by the non--QCD--like
dynamics of technicolor with a slowly running gauge coupling, walking
technicolor, which will be described in the next section.

\subsection{\it {\underbar{Precision Electroweak Measurements}}}

Precision electroweak measurements actually challenge technicolor, not
extended technicolor. The basic parameters of the standard $SU(2)\otimes
U(1)$ model---$\alpha(M_Z)$, $M_Z$, $\sin^2 \theta_W$---are measured so
precisely that they may be used to limit new physics at energy scales above
100~GeV~\cite{pettests}. The quantities most sensitive to new physics are
defined in terms of correlation functions of the electroweak currents:
\be\label{eq:pifcn}
\int d^4x \ts e^{-i q\cdot x} \langle\Omega | T\left(j^\mu_i(x)
j^\nu_j(0)\right) | \Omega \rangle =
i g^{\mu\nu} \Pi_{ij}(q^2) + q^\mu q^\nu \ts {\rm terms} \ts.
\ee
Once one has accounted for the contributions from standard model physics,
including a single Higgs boson (whose mass $M_H$ must be assumed), new
high--mass physics affects the $\Pi_{ij}$ functions. Assuming that the scale
of this physics is well above $M_{W,Z}$, it enters the ``oblique'' correction
factors $S$, $T$, $U$ defined by
\bea\label{eq:stu}
&&S= 16\pi {d \over {d q^2}} \left[ \Pi_{33} (q^2) - \Pi_{3Q}(q^2)
\right]_{q^2=0} \ts \equiv \ts 16\pi \left[ \Pi_{33}^{'}(0) - \Pi_{3Q}^{'}(0)
\right] \ts , \nn\\
&&T= {4\pi \over{M^2_Z \cos^2\theta_W \sin^2\theta_W}}
\ts \left[ \Pi_{11}(0) - \Pi_{33}(0) \right] \ts , \nn\\
&&U= 16\pi \left[ \Pi_{11}^{'}(0) - \Pi_{33}^{'}(0) 
\right] \ts .
\eea
The parameter $S$ is a measure of the splitting between $M_W$ and $M_Z$
induced by weak--isospin conserving effects; the $\rho$--parameter is given
by $\rho \equiv M^2_W/M^2_Z \cos^2\thw = 1 + \alpha T$; the $U$--parameter
measures weak--isospin breaking in the $W$ and $Z$ mass splitting. The
experimental limits on $S,T,U$ are~\cite{pdg}
\bea\label{eq:stuvalues}
&&S= -0.07 \pm 0.11 \ts\ts (-0.09)\ts,\nn\\
&&T= -0.10 \pm 0.14 \ts\ts (+0.09)\ts,\nn\\
&&U= +0.11\pm 0.15  \ts\ts (+0.01)\ts.
\eea
The central values assume $M_H = 100\,\gev$, and the parentheses contain the
change for $M_H = 300\,\gev$. The $S$ and $T$--parameters and $M_H$ cannot be
obtained simultaneously from data because the Higgs loops resemble oblique
effects.

The $S$--parameter is the one most touted as a show--stopper for
technicolor \cite{pettests,tctests}. The value obtained in technicolor by
scaling from QCD is $\CO(1)$. For example, for $N$ color--singlet
technidoublets, Peskin and Takeuchi found the positive result
\be\label{eq:svalue}
S = 4 \pi \left(1 + {M^2_{\rho_T} \over{M^2_{a_{1T}}}}\right ) {F^2_\pi \over
{M^2_{\rho_T}}} \simeq 0.25 \ts N {N_{TC}\over{3}} \ts.
\ee
The resolution to this problem may also be found in walking technicolor. One
thing is sure: naive scaling of $S$ from QCD is unjustified and probably
incorrect in walking gauge theories. No reliable estimate exists because no
data on walking gauge theories are available to put into the calculation of
$S$.

\subsection{\it {\underbar{The Top Quark Mass}}}

The ETC scale required to produce $m_t = 175\,\gev$ in Eq.~(3) is
$0.75\,\tev/N^{3/4}$ for $N$ technidoublets. This is uncomfortably close to
the TC scale itself. In effect, TC becomes strong and ETC is broken at
the same energy; the representation of broken ETC interactions as contact
operators is wrong; and all our mass estimates are questionable. It is
possible to raise the ETC scale so that it is considerably greater than
$m_t$, but then one runs into the problem of fine--tuning the ETC coupling
$g_{ETC}$ (just as in the Nambu--Jona-Lasinio model, where requiring the
dynamical fermion mass to be much less than the four--fermion mass scale
$\Lambda$ requires fine--tuning the NJL coupling very close to
$4\pi$)~\cite{setc}. This flouts our cherished principle of naturalness, and
we reject it. Another, more direct, problem with ETC generation of the top
mass is that there must be large weak isospin violation to raise it so high
above the bottom mass. This adversely affects the $\rho$
parameter~\cite{tombowick}. The large effective ETC coupling to top quarks
also makes a large, unwanted contribution to the $Z \ra \ol b b$ decay rate,
in conflict with experiment~\cite{zbbth}.

In the end, there is no plausible way to understand the top quark's large
mass from ETC. Something more is needed. The best idea so far is
topcolor--assisted technicolor~\cite{tctwohill}, in which a new gauge
interaction, topcolor~\cite{topcref}, becomes strong near 1~TeV and generates
a large $\ol t t$ condensate and top mass. This, too, will be described in
the next section.

\section{Dynamical Rescues}

The FCNC and STU difficulties of technicolor have a common cause: the
assumption that technicolor is a just a scaled--up version of QCD. Let us
focus on Eqs.(3,4,6), the key equations of extended technicolor. In a
QCD--like technicolor theory, asymptotic freedom sets in quickly above
$\LTC$, the anomalous dimension $\gamma_m \ll 1$, and $\condetc \simeq
\condtc$. The conclusion that fermion and technipion masses are one or more
orders of magnitude too small then followed from the FCNC requirement in
Eqs.~(\ref{eq:dmlimit},\ref{eq:epslimit}) that $\METC/g_{ETC} |V_{ds}| \simge
1000\,\tev$. Scaling from QCD also means that the technihadron spectrum is
just a magnified image of the QCD--hadron spectrum, hence that $S$ is too
large for all technicolor models except, possibly, the minimal one--doublet
model with $N_{TC} \simle 4$.

The solution to these difficulties lies in technicolor gauge dynamics that are
distinctly not QCD--like. The only plausible example is one in which
the gauge coupling $\atc(\mu)$ evolves slowly, or ``walks'', over the large
range of energy $\LTC \simle \mu \simle \METC$~\cite{wtc}. In the extreme
walking limit in which $\atc(\mu)$ is constant, it is possible to obtain an
approximate nonperturbative formula for the $\ol T T$ anomalous dimension
$\gamma_m$, namely,
\be\label{eq:nonpert}
\gamma_m(\mu) = 1 - \sqrt{1-\atc(\mu)/\alpha_C} \quad {\rm where} \ts\ts\ts 
\alpha_C  = {\pi \over{3 C_2(R)}} \ts.
\ee
This reduces to the expression in Eq.~(\ref{eq:gmm}) for small $\atc$.  It
has been argued that $\gamma_m = 1$ is the signal for spontaneous chiral
symmetry breaking~\cite{agchmg}, and, so, $\alpha_C$ is called the critical
coupling for $\chi$SB, with $\pi/3C_2(R)$ its approximate value.\footnote{An
attempt to improve upon this approximation and study its accuracy is in
Ref.~\cite{alm}.} If we identify $\LTC$ with the scale at which
technifermions in the $\sutc$ fundamental representation condense, then
$\atc(\LTC) = \alpha_C$.

In walking technicolor, $\atc(\mu)$ is presumed to remain close to its
critical value from $\LTC$ almost up to $\METC$. This implies $\gamma_m(\mu)
\simeq 1$, and by Eq.~(\ref{eq:condrenorm}), the condensate $\condetc$ is
enhanced by a factor of 100 or more. This yields quark masses up to a few GeV
and reasonably large technipion masses despite the very large ETC mass
scale. This is still not enough to account for the top mass; more on that
soon.

Another consequence of the walking $\atc$ is that the spectrum of
technihadrons, especially the $I=0,1$ vector and axial vector mesons, $\tro$,
$\tom$, $a_{1T}$ and $f_{1T}$, cannot be
QCD--like~\cite{kltasi,klglasgow,edrta}. In QCD, the lowest lying isovector
$\rho$ and $a_1$ saturate the spectral functions appearing in Weinberg's sum
rules~\cite{sfsr}. Then, the relevant combination $\rho_V - \rho_A$ of
spectral functions falls off like $1/p^6$ for $p > M_{\rho,a_1} \sim
\Lambda_{QCD}$, and the spectral integrals converge very rapidly. This
``vector meson dominance'' of the spectral integrals is related to the
precocious onset of asymptotic freedom in QCD. The $1/p^6$ momentum
dependence is just what one would deduce from a naive, lowest--order
calculation of $\rho_V - \rho_A$ using the asymptotic $1/p^2$ behavior of the
quark dynamical mass $\Sigma(p)$~\cite{kdlhdp}. In walking technicolor, the
technifermion's $\Sigma(p)$ falls only like $1/p^{(2-\gamma_m)} \sim 1/p$ for
$\LTC \simle \METC$, so that $\rho_V - \rho_A \sim 1/p^4$ up to very high
energies. To account for this in terms of spin--one technihadrons, there must
be something like a tower of $\tro$ and $\tom$ extending up to $\METC$. Their
mass spectrum, widths, and couplings to currents cannot be predicted. Thus,
without experimental knowledge of these states, it is impossible to
estimate $S$ reliably, any more than it would have been in QCD before the
$\rho$ and $a_1$ were discovered and measured.

Another issue that may affect $S$ is that it is usually
defined assuming that the new physics appears at energies well above
$M_{W,Z}$. We shall see below that, on the contrary, walking technicolor
suggests that there are $\tpi $ and $\tro$ starting near or not far above
$100\,\gev$.

We have seen that extended technicolor cannot explain the top quark's large
mass. An alternative approach was developed in the early 90s based on a new
interaction of the third generation quarks.  This interaction, called
topcolor, was invented as a minimal dynamical scheme to reproduce the
simplicity of the one--doublet Higgs model {\it and} explain a very large
top--quark mass~\cite{topcref}. Here, a large top--quark condensate,
$\condtbt$, is formed by strong interactions at the energy scale,
$\Lambda_t$~\cite{topcondref}. To preserve electroweak $SU(2)$, topcolor must
treat $t_L$ and $b_L$ the same. To prevent a large $b$--condensate and mass,
it must violate weak isospin and treat $t_R$ and $b_R$ differently. In order
that the resulting low--energy theory simulate the standard model,
particularly its small violation of weak isospin, the topcolor scale must be
very high---$\Lambda_t \sim 10^{15}\,\gev \gg m_t$. Therefore, this original
topcolor scenario is highly unnatural, requiring a fine--tuning of couplings
of order one part in $\Lambda_t^2/m_t^2 \simeq 10^{25}$ (remember
Nambu--Jona-Lasinio!).

Technicolor is still the most natural mechanism for electroweak symmetry
breaking, while topcolor dynamics most aptly explains the top mass. Hill
proposed to combine the two into what he called topcolor--assisted
technicolor (TC2)~\cite{tctwohill}. In TC2, electroweak symmetry breaking is
driven mainly by technicolor interactions strong near $1\,\tev$. Light quark,
lepton, and technipion masses are still generated by ETC. The topcolor
interaction, whose scale is also near $1\,\tev$, generate $\condtbt$ and the
large top--quark mass.\footnote{Three massless Goldstone ``top--pions''
arise from top-quark condensation. Thus, ETC interactions must contribute a
few~GeV to $m_t$ to give the top--pions a mass large enough that $t \ra b
\pi^+_t$ is not a major decay mode.} The scale of ETC interactions still must
be at least several $100\,\tev$ to suppress flavor-changing neutral currents
and, so, the technicolor coupling still must walk. Their marriage neatly
removes the objections that topcolor is unnatural and that technicolor cannot
generate a large top mass. In this scenario, the nonabelian part of topcolor
is an ordinary asymptotically free gauge theory.

Hill's original TC2 scheme assumes separate color $SU(3)$ and weak
hypercharge $U(1)$ gauge interactions for the third and for the first two
generations of quarks and leptons. In the simplest example, the (electroweak
eigenstate) third generation $(t,b)_{L,R}$ transform with the usual quantum
numbers under the topcolor gauge group $\suone \otimes \uone$ while $(u,d)$,
$(c,s)$ transform under a separate group $\sutwo \otimes \utwo$.  Leptons of
the third and the first two generations transform in the obvious way to
cancel gauge anomalies. At a scale of order $1\,\tev$, $\suone \otimes \sutwo
\otimes \uone \otimes \utwo$ is dynamically broken to the diagonal subgroup
of ordinary color and weak hypercharge, $SU(3)_C \otimes \uy$. At this
energy, the $\suone \otimes \uone$ couplings are strong while the $\sutwo
\otimes \utwo$ couplings are weak. This breaking gives rise to massive gauge
bosons---a color octet of ``colorons'' $V_8$ and a color singlet $Z'$.

Top, but not bottom, condensation is driven by the fact that the $\suone
\otimes \uone$ interactions are supercritical for top quarks, but subcritical
for bottom.\footnote{A large bottom condensate is not generated by $\suone$
because it is broken and its coupling does not grow stronger as one descends
to lower energies.} The difference between top and bottom is caused by the
$\uone$ couplings of $t_R$ and $b_R$. If this TC2 scenario is to be natural,
i.e., there is no fine--tuning of the $\suone$, the $\uone$ couplings {\it
cannot} be weak. To avoid large violations of weak isospin in this and all
other TC2 models~\cite{cdt}, right as well as left--handed members of
individual technifermion doublets $T_{L,R} = (T_U, T_D)_{L,R}$ must carry the
same $\uone$ quantum quantum numbers, $Y_{1L}$ and $Y_{1R}$,
respectively~\cite{tctwoklee}.

Hill's simplest TC2 model does not how explain how topcolor is broken. Since
natural topcolor requires it to occur near 1~TeV, the most likely cause is
technifermion condensation. In Ref.~\cite{tctwokl}, it was argued that this
can be done for $\suone\otimes\sutwo\ra SU(3)_C$ by arranging that
technifermion doublets $T_1$ and $T_2$ transforming under $\sutc \otimes
\suone \otimes \sutwo$ as $(\Ntc,3,1)$ and $(\Ntc,1,3)$ condense with each
other as well as themselves, i.e.,
\be\label{eq:tconds} \langle \ol T_{iL} T_{jR} \rangle = -U_{ij} \Delta_T
\quad (i,j = 1,2)\ts,
\ee
where $U$ is a nondiagonal unitary matrix and $\Delta_T$ the technifermion
condensate of $\CO(\LTC^3)$. The strongly coupled $\uone$ plays a critical
role in tilting $U$ away from the identity, which is the form of the
condensate preferred by the color interactions.

The breaking $\uone\otimes\utwo \ra \uy$ is trickier. In order that there is
a well--defined $U(1)_Y$ boson with standard couplings to all quarks and
leptons, this must occur at a somewhat higher scale, several TeV.  Thus, the
$Z'$ boson from this breaking has a mass of several TeV and is {\it strongly}
coupled to technifermions, at least.\footnote{In Ref.~\cite{tctwokl} the
fermions of the first two generations also need to couple to $\uone$. The
limits on these strong couplings and $M_{Z'}$ from precision electroweak
measurements were studied by Chivukula and Terning~\cite{rscjt}. Another
variant of TC2 has all three generations transforming in the same way under
topcolor~\cite{ccs}. This ``flavor--universal topcolor'' has certain
phenomenological advantages (see the second paper of Ref.~\cite{tctwokl}),
but the problems of the strong $\uone$ coupling afflict it too.} To employ
technicolor in this $U(1)$ breaking too, technifermions $\psi_{L,R}$
belonging to a higher--dimensional $\sutc$ representation are
introduced. They condense at higher energy than the fundamentals
$T_{iL,R}$~\cite{multiklee}. The critical reader will note that this scenario
also flirts with unnatural fine tuning because the multi--TeV $Z'$ plays a
critical role in top and bottom quark condensation. Another pitfall is that
the strong $\uone$ coupling may blow up at a Landau singularity at a
relatively low energy~\cite{tctwokl,rsctriv}. To avoid this, unification of
$\uone$ with the nonabelian $G_{ETC}$ must occur at a lower energy
still. This is not a very satisfactory state of affairs, but that is how
things stand for now with TC2. There are many opportunities for improvement.

A variant of topcolor models is called the ``top seesaw''
mechanism~\cite{seesaw}. Its motivation is to realize the original,
supposedly more economical, top--condensate idea of the Higgs boson as a
fermion--antifermion bound state~\cite{topcondref}. Apart from its fine
tuning problem, that way failed because it implied a top mass of about
250~GeV. In top seesaw models, an electroweak singlet fermion $F$ acquires a
dynamical mass of {\it several} TeV. Through mixing of $F$ with the top
quark, it gives the latter a much smaller mass (the seesaw) and the scalar
$\ol F F$ bound state acquires a component with an electroweak symmetry
breaking vacuum expectation value. The latest twist on this variant is called
the ``topcolor jungle gym''~\cite{jungle}. We'll say no more about these
approaches here as they are off our main line of technicolor and extended
technicolor. The interested reader should consult the literature.

{\begin{table}{
\centering
\begin{tabular}{|c|c|c|}
\hline
$V_T$ Decay Mode& 
$V(V_T \ra G\tpi) \times M_V/e$ & 
$A(V_T \ra G\tpi) \times M_A/e$  
\\
\hline\hline
$\tom \ra \gamma \tpiz$& $\cos\chi$ & 0 \\
$\ts\ts\ts\quad \ra \gamma \tpipr$ & $(Q_U + Q_D)\ts \cos\chi'$ & 0 \\ 
$\qquad \ra Z^0 \tpiz$ & $\cos\chi\cot 2\thw$ & 0 \\ 
$\ts\qquad \ra Z^0 \tpipr$ & $-(Q_U+Q_D)\ts \cos\chi'\tan \thw$ & 0 \\ 
$\ts\ts\ts\ts\qquad \ra W^\pm \tpimp$ & $\cos\chi/(2\sin\thw)$ & 0 \\ 
\hline
$\troz \ra \gamma \tpiz$ & $(Q_U + Q_D)\ts \cos\chi$ & 0 \\
$\ts\ts\ts\quad \ra \gamma \tpipr$ & $\cos\chi'$ & 0 \\
$\qquad \ra Z^0 \tpiz$ & $-(Q_U+Q_D)\ts \cos\chi \tan \thw$ & 0 \\
$\ts\qquad \ra Z^0 \tpipr$ & $\cos\chi'\ts \cot 2\thw$ & 0 \\
$\ts\ts\ts\ts\qquad \ra W^\pm \tpimp$ & 0 & $-\cos\chi/(2\sin\thw)$ \\
\hline
$\tropm \ra \gamma \tpipm$ & $(Q_U + Q_D)\ts \cos\chi$ & 0 \\ 
$\qquad \ra Z^0 \tpipm$ & $-(Q_U+Q_D)\ts \cos\chi \tan \thw$ & $\cos\chi
\ts /\sin 2\thw$ \\  
$\ts\ts\ts\qquad \ra W^\pm \tpiz$ & 0 & $\cos\chi/(2\sin\thw)$ \\ 
$\ts\ts\ts\qquad \ra W^\pm \tpipr$ & $\cos\chi'/(2\sin\thw)$ & 0 \\
\hline\hline
\end{tabular}}
\medskip
\caption{{\it Relative vector and axial vector amplitudes for $V_T \ra G
\tpi$ with $V_T = \tro,\tom$ and $G$ a transverse electroweak boson,
$\gamma,Z^0,W^\pm$; from Ref.~\cite{tcsm_singlet}.}}
%
\end{table}}
\begin{figure}[t]
 \vspace{9.0cm}
\includegraphics{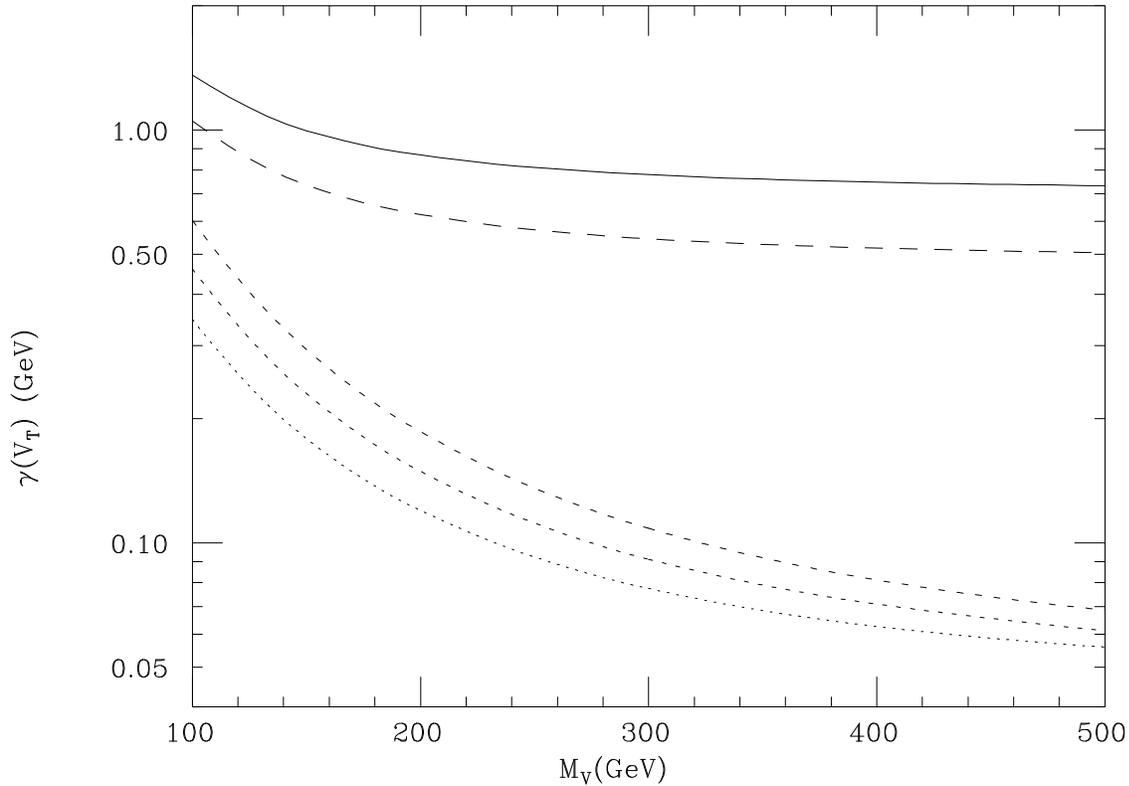}
\vskip2.5truecm
 \caption{\it
      Technivector meson total decay rates versus $M_V = M_A$ for $\troz$
  (solid curve) and $\tropm$ (long-dashed) with $M_{\tro} = 210\,\gev$, and
  $\tom$ with $M_{\tom} = 200$ (lower dotted), 210 (lower short-dashed), and
  $220\,\gev$ (lower medium-dashed); $Q_U + Q_D = 5/3$ and $M_{\tpi} =
  110\,\gev$; from Ref.~\cite{tcsm_singlet}.
    \label{fig2} }
\end{figure}
\begin{figure}[t]
 \vspace{9.0cm}
\includegraphics{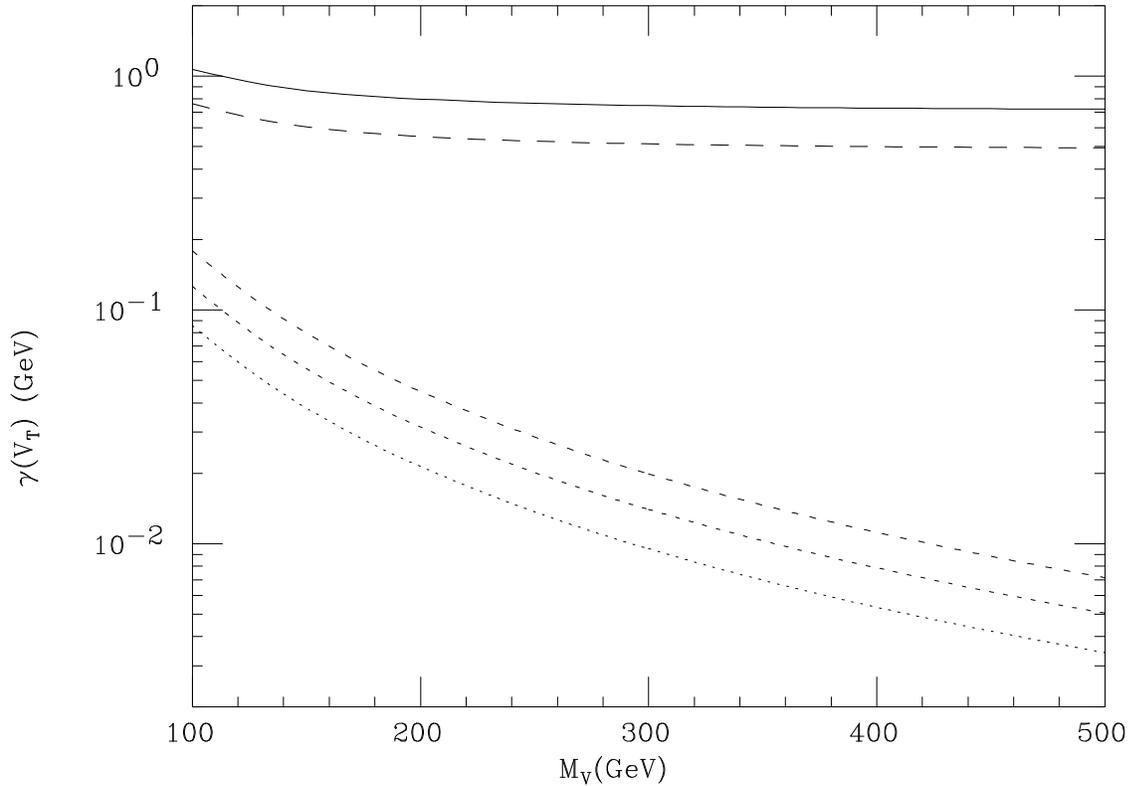}
\vskip2.5truecm
 \caption{\it
      Decay rates as in Fig.~2, with $Q_U + Q_D = 0$; from
      Ref.~\cite{tcsm_singlet}.
    \label{fig3} }
\end{figure}

\begin{figure}[t]
 \vspace{9.0cm}
\includegraphics{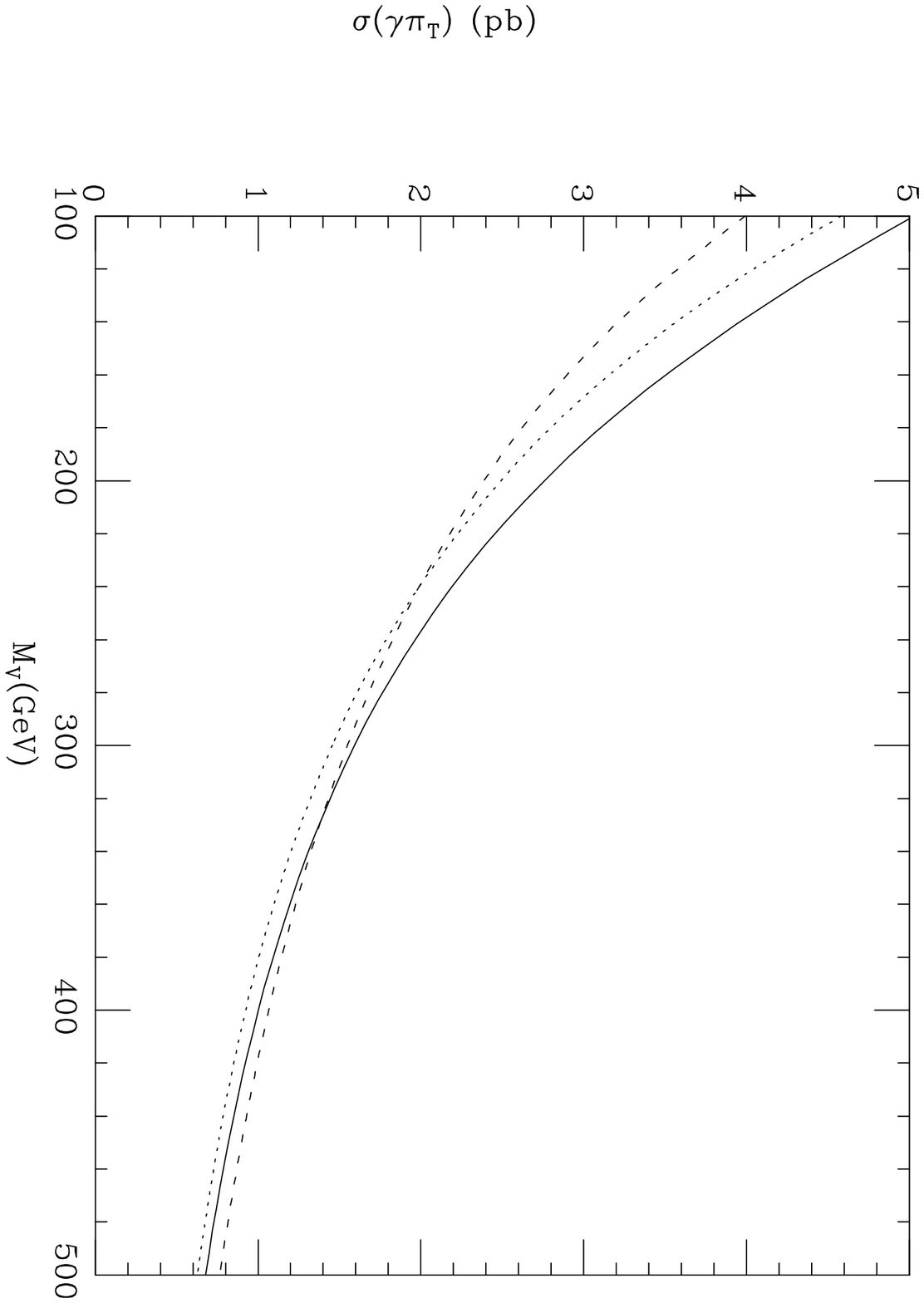}
\vskip2.5truecm
 \caption{\it
   Production rates in $\ol p p$ collisions at $\ecm = 2\,\tev$
   for the sum of $\tom$, $\troz$, $\tropm \ra \gamma
   \tpi$ versus $M_V$, for $M_{\tro} = 210\,\gev$ and $M_{\tom} = 200$ (dotted
   curve), 210 (solid), and $220\,\gev$ (short-dashed); $Q_U + Q_D = 5/3$, and
   $M_{\tpi} = 110\,\gev$; from Ref.~\cite{tcsm_singlet}.
   \label{fig4} }
\end{figure}
\begin{figure}[t]
 \vspace{9.0cm}
\includegraphics{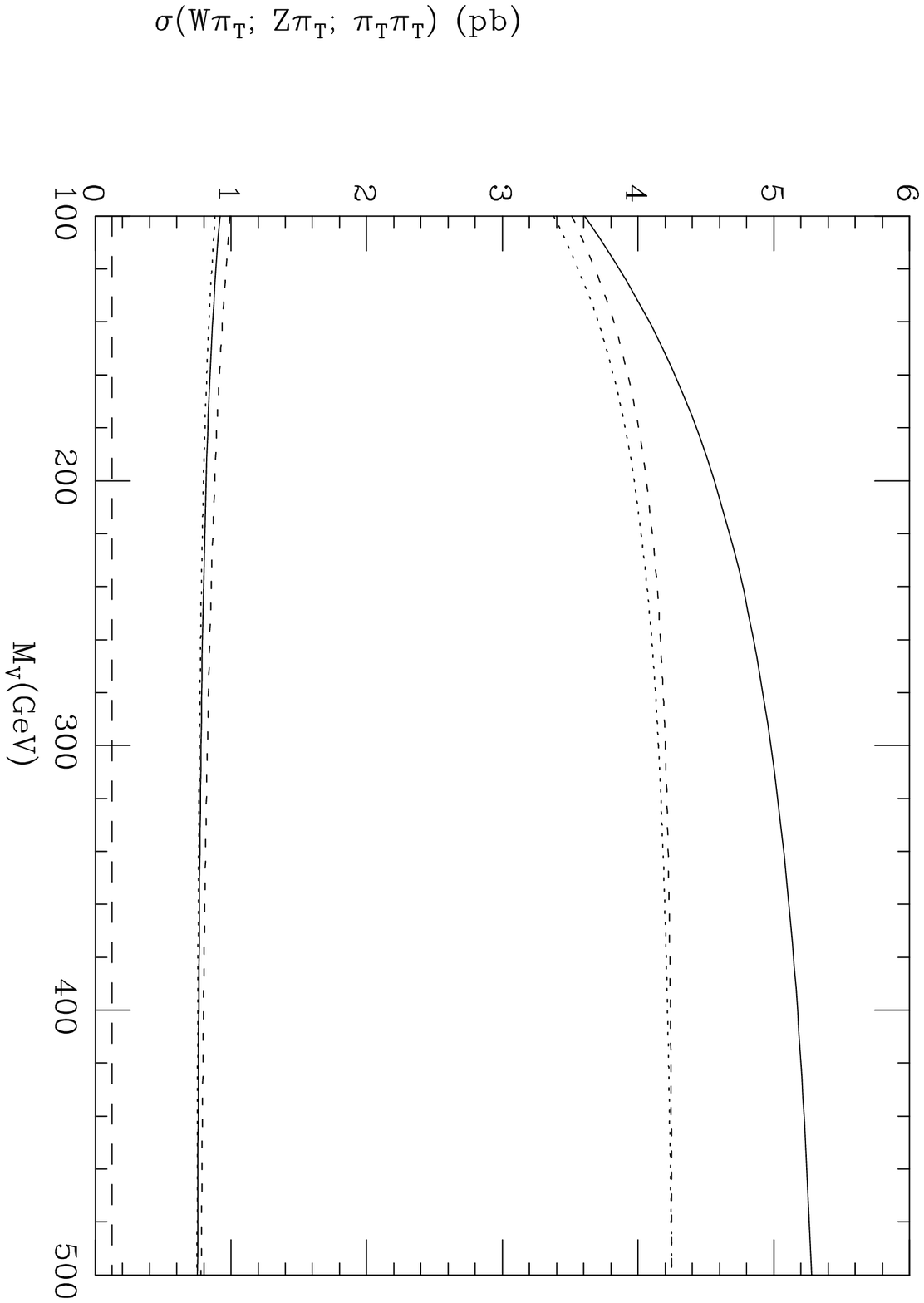}
\vskip2.5truecm
 \caption{\it
   Production rates in $\ol p p$ collisions at $\ecm = 2\,\tev$ for $\tom$,
   $\troz$, $\tropm \ra W \tpi$ (upper curves) and $Z\tpi$ (lower curves)
   versus $M_V$, for $M_{\tro} = 210\,\gev$ and $M_{\tom} = 200$ (dotted
   curve), 210 (solid), and $220\,\gev$ (short-dashed); $Q_U + Q_D = 5/3$ and
   $M_{\tpi} = 110\,\gev$. Also shown is $\sigma(\tro \ra \tpi\tpi)$ (lowest
   dashed curve); from Ref.~\cite{tcsm_singlet}.
    \label{fig5} }
\end{figure}

\begin{figure}[t]
 \vspace{9.0cm}
\includegraphics{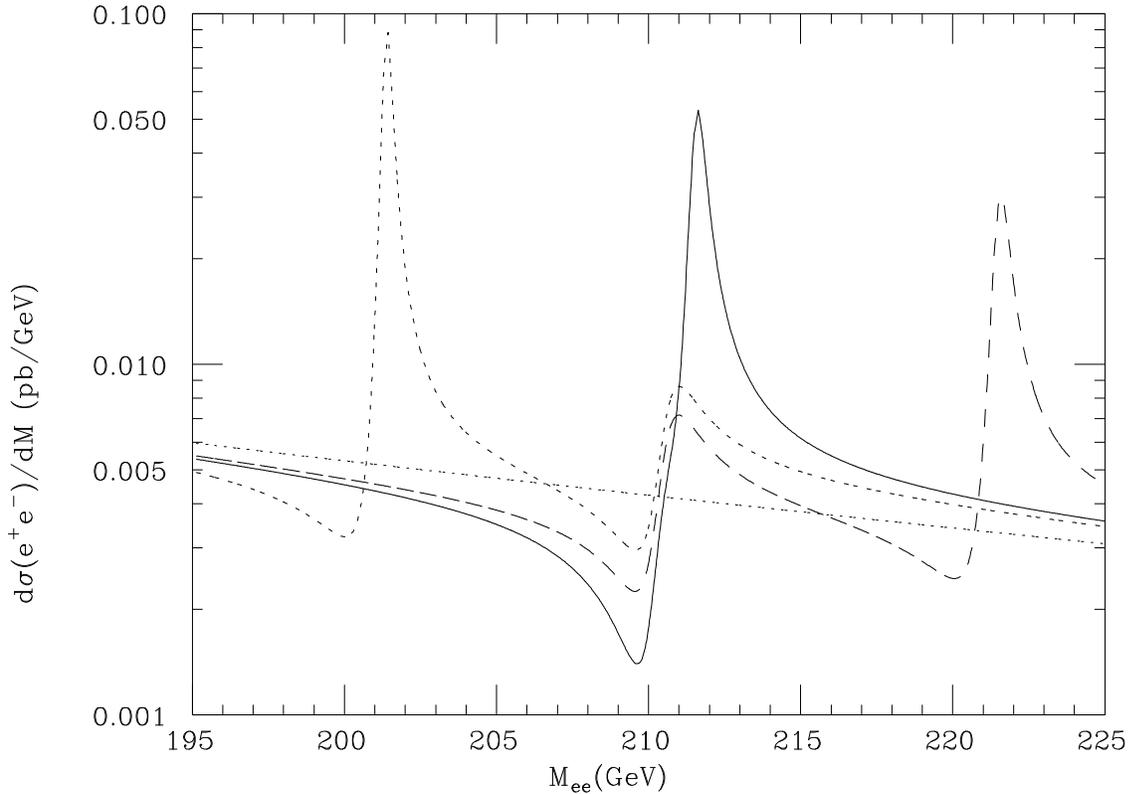}
\vskip2.5truecm
 \caption{\it
   Invariant mass distributions in $\ol p p$ collisions at $\ecm = 2\,\tev$
for $\tom$, $\troz \ra e^+e^-$ for $M_{\tro} = 210\,\gev$ and $M_{\tom} =
200$ (short-dashed curve), 210 (solid), and $220\,\gev$ (long-dashed); $M_V =
100\,\gev$. The standard model background is the sloping dotted line. $Q_U +
Q_D = 5/3$ and $M_{\tpi} = 110\,\gev$; from Ref.~\cite{tcsm_singlet}.
    \label{fig6} }
\end{figure}
\begin{figure}[t]
 \vspace{9.0cm}
\includegraphics{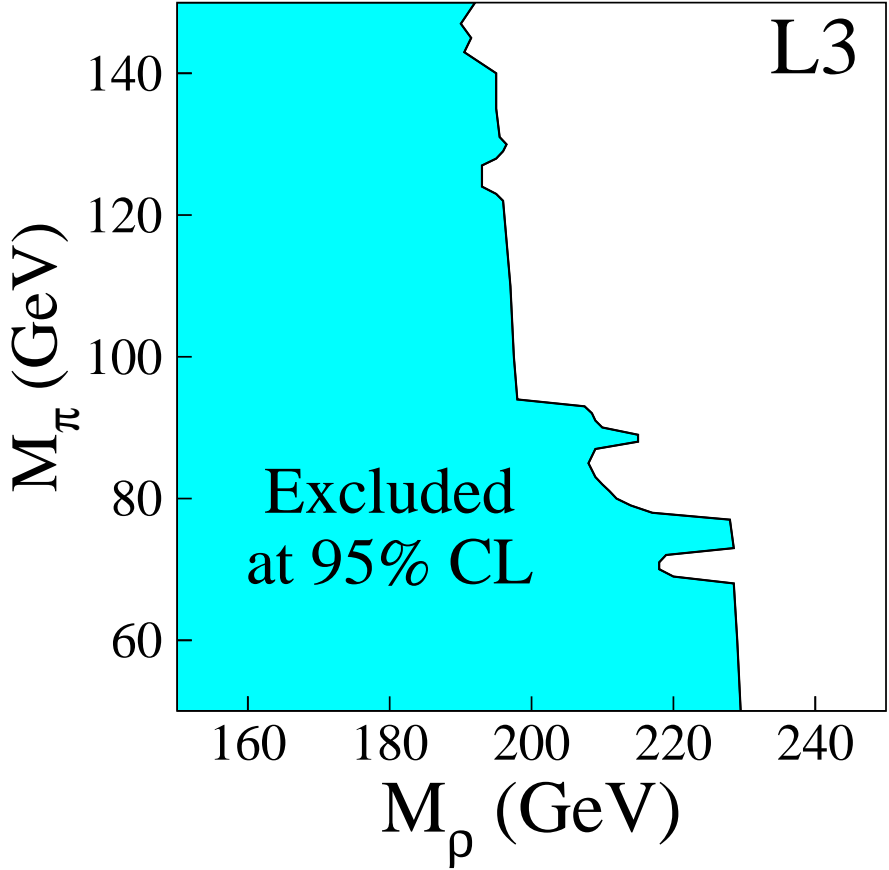}
\vskip3.0truecm
 \caption{\it
      The $M_{\tro}$--$M_{\tpi}$ region excluded by L3 at the 95\% CL; from
      Ref.~\cite{L3}.
    \label{fig7} }
\end{figure}
\begin{figure}[t]
 \vspace{9.0cm}
\includegraphics{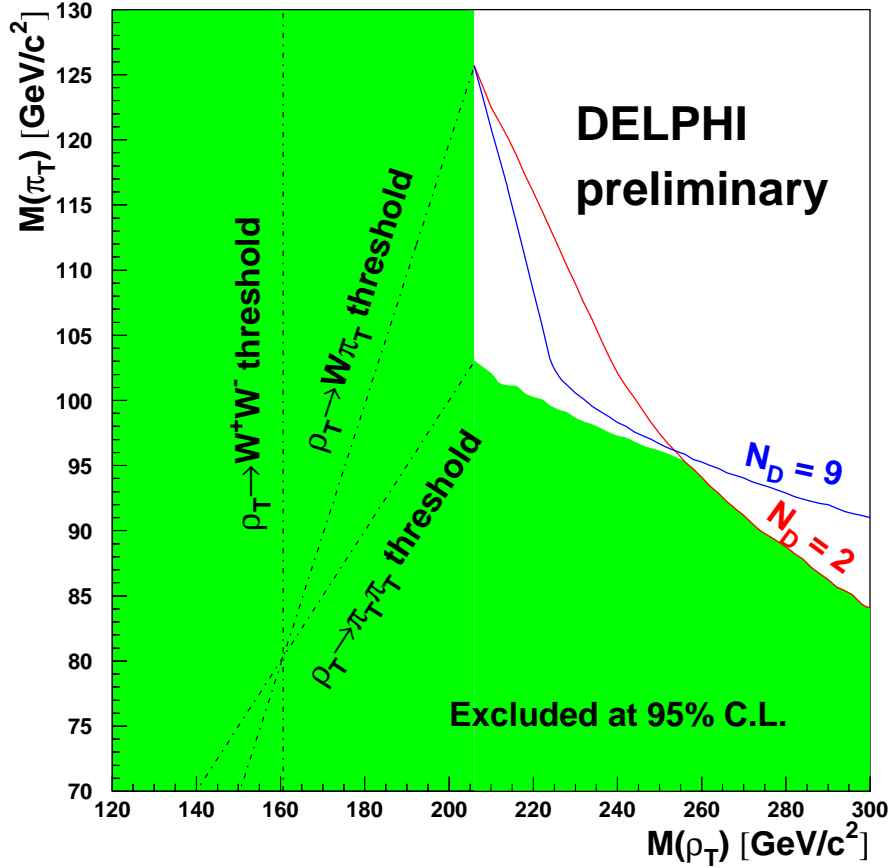}
\vskip3.0truecm
 \caption{\it
      The $M_{\tro}$--$M_{\tpi}$ region excluded at the 95\% CL by the DELPHI
      analysis of Ref.~\cite{delphi}.
    \label{fig8} }
\end{figure}
\begin{figure}[t]
 \vspace{9.0cm}
\includegraphics{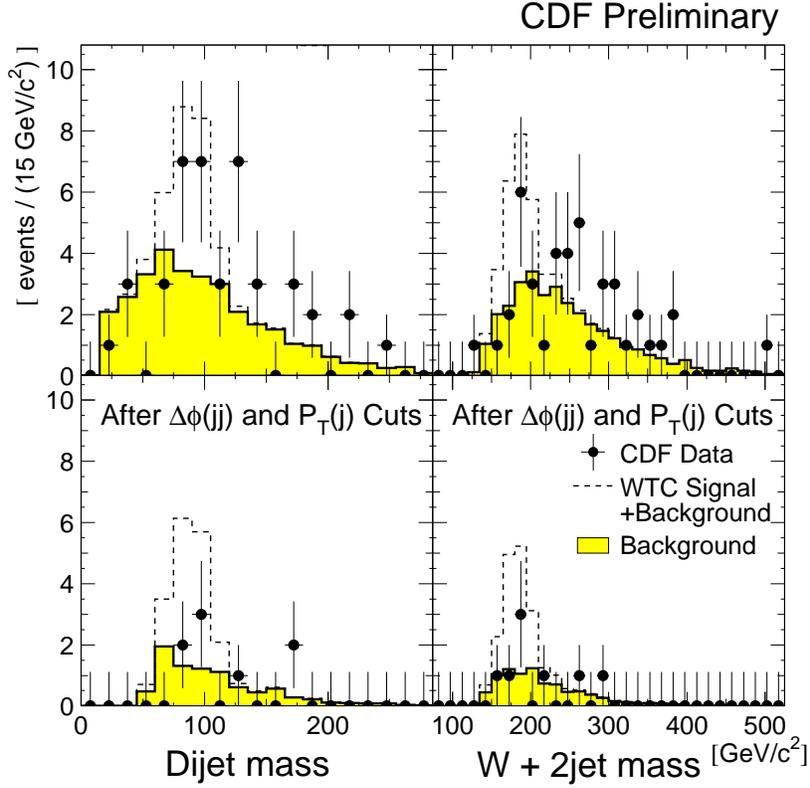}
\vskip2.3truecm
 \caption{\it
      Invariant mass of the dijet system and of the $W+2\ts\jet$ system for
      the $\ell+2\ts\jet$ mode; from Ref.~\cite{cdfwpi}. The mass combination
      shown is $M_{\tpi} =90\,\gev$ and $M_{\tro}=180\,\gev$.
    \label{fig9} }
\end{figure}
\begin{figure}[t]
 \vspace{9.0cm}
\includegraphics{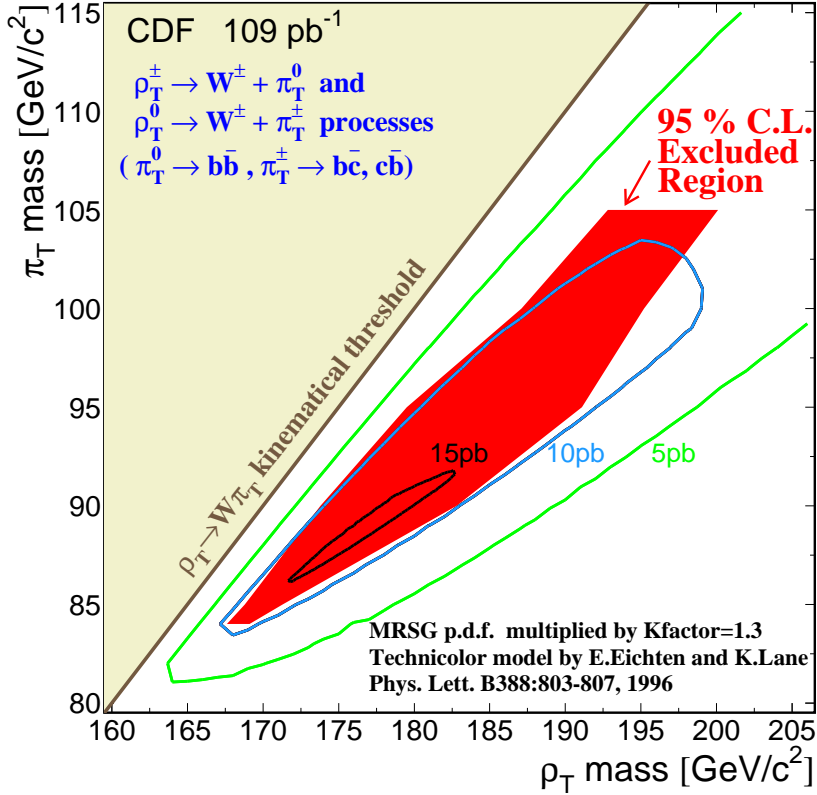}
\vskip2.3truecm
 \caption{\it
   Excluded region for the CDF search for $\tro \ra W^\pm \tpi$ in
   Ref.~\cite{cdfwpi}. See that reference for an explanation of the 5, 10,
   15~pb contours.
    \label{fig10} }
\end{figure}
\begin{figure}[t]
 \vspace{9.0cm}
\includegraphics{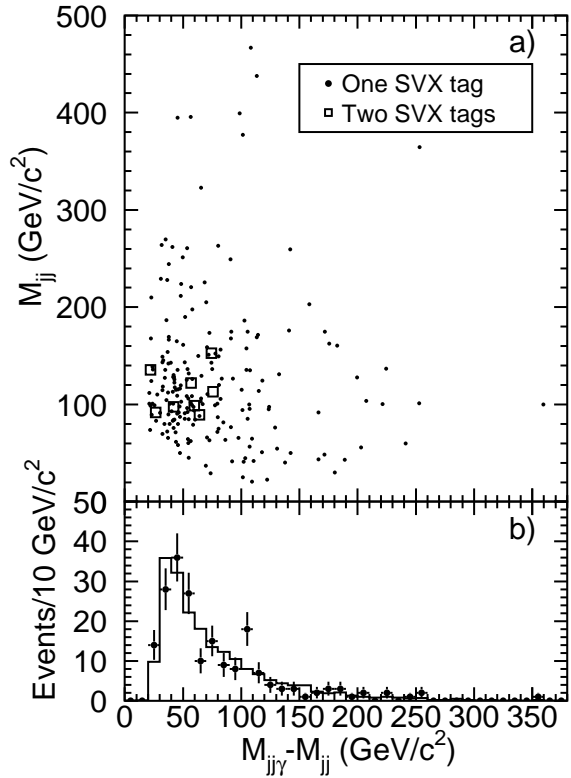}
\vskip2.0truecm
 \caption{\it
      (a) The distribution of $M_{jj}$ vs. $M_{jj\gamma} - M_{jj}$ for events
      with a photon, $b$--tagged jet and a second jet. (b) Projection of this
      data in $M_{jj\gamma} - M_{jj}$; from Ref.~\cite{cdfgpi}.
    \label{fig11} }
\end{figure}
\begin{figure}[t]
\vspace{9.0cm}
\includegraphics{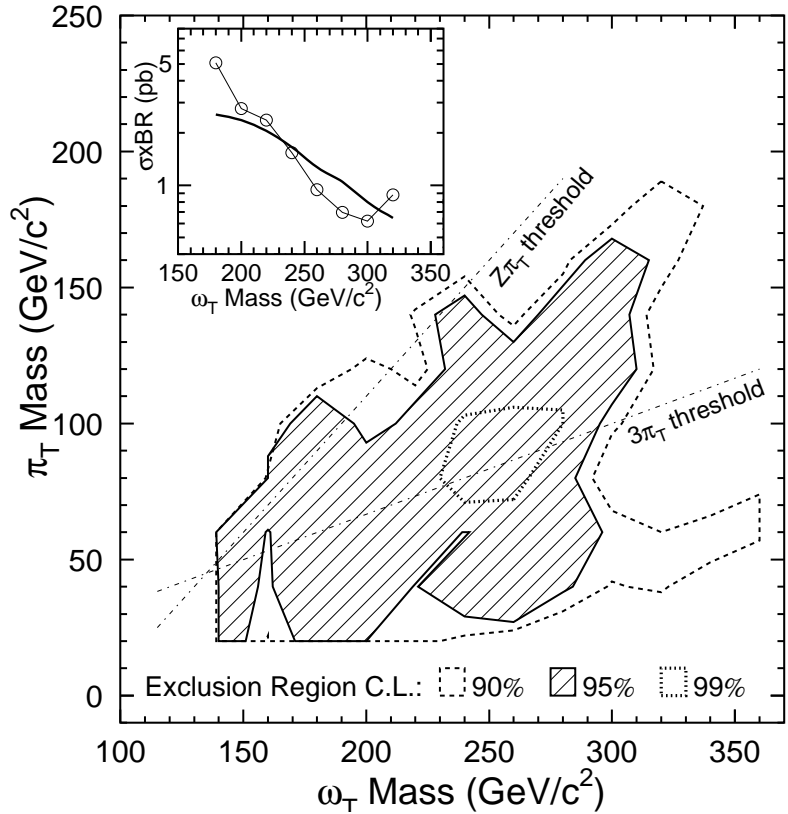}
\vskip2.5truecm
\caption{\it
  The 90\%, 95\% and 99\% CL exclusion regions for the CDF search for $\tom
  \ra \gamma\tpi$ in Ref.~\cite{cdfgpi}. The inset shows the limit on $\sigma
  B$ for $M_{\tpi} = 120\,\gev$. The circles represent the limit and the
  solid line the prediction from the second paper in Ref.~\cite{elw}.
  \label{fig12} }
\end{figure}
\begin{figure}[t]
\vspace{9.0cm}
\includegraphics{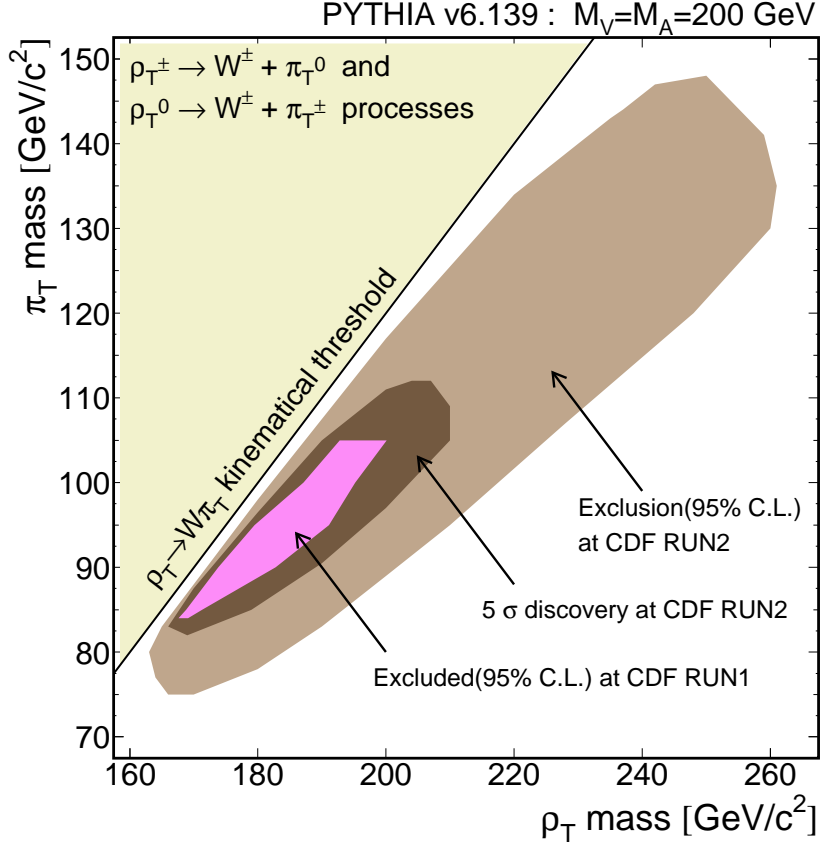}
\vskip2.7truecm
\caption{\it
  Reach of the CDF detector in Tevatron Run~IIa for $\tro \ra W^\pm \tpi$
  with $M_V = M_A = 200\,\gev$; from Ref.~\cite{handa}.
\label{fig13} }
\end{figure}

\begin{figure}[t]
\vspace{9.0cm}
\includegraphics{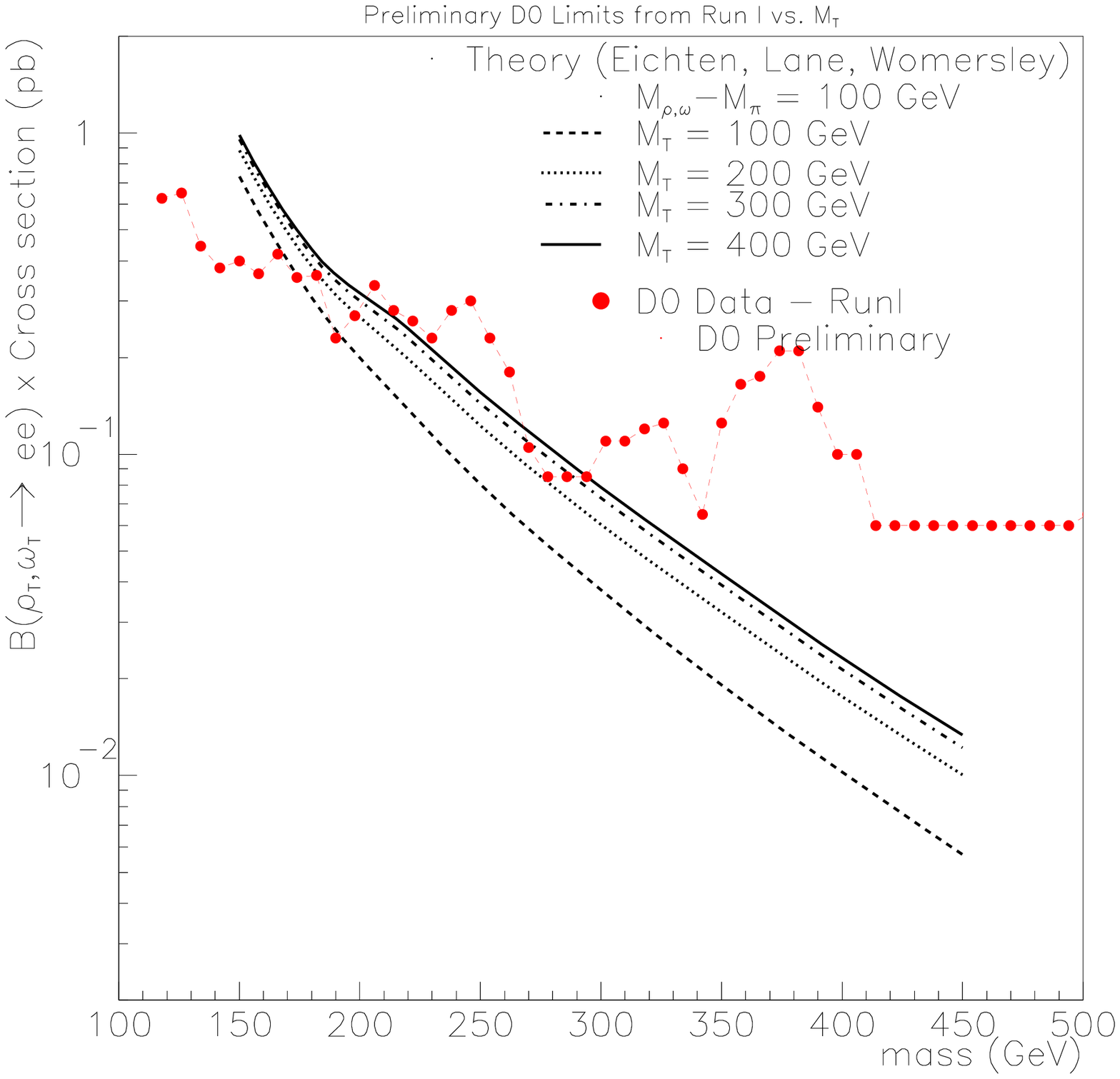}
\vskip1.5truecm
\caption{\it
  Excluded regions for the D\O\ search for $\troz,\tom \ra e^+e^-$; from
  Ref.~\cite{dzeroee}.
\label{fig14} }
\end{figure}
\begin{figure}[t]
 \vspace{9.0cm}
\includegraphics{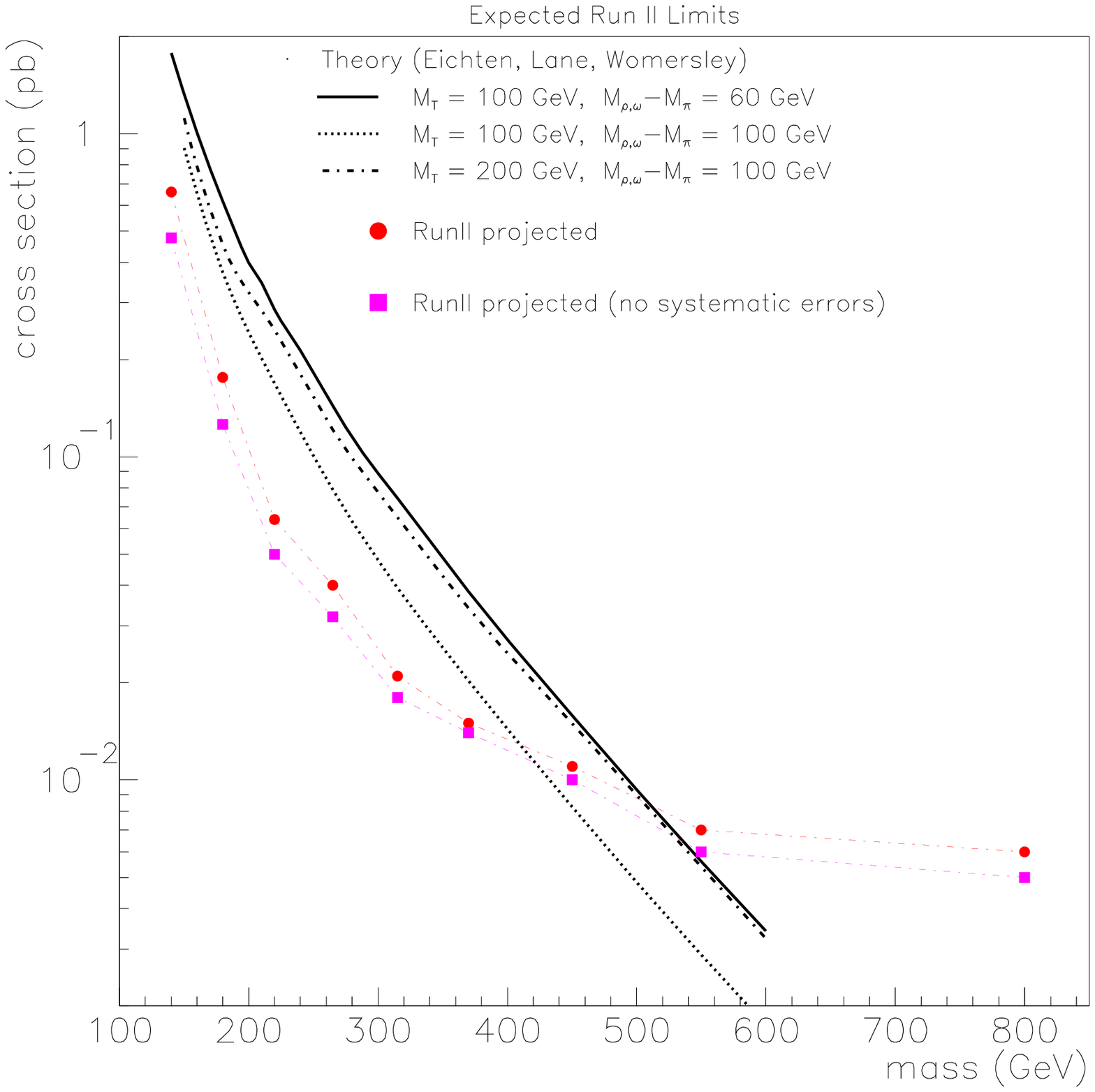}
\vskip2.0truecm
 \caption{\it
      Reach of the D\O\ detector in Tevatron Run~IIa for $\troz,\tom \ra
      e^+e^-$; from Ref.~\cite{meena}.
    \label{fig15} }
\end{figure}
\begin{figure}[t]
 \vspace{9.0cm}
\includegraphics{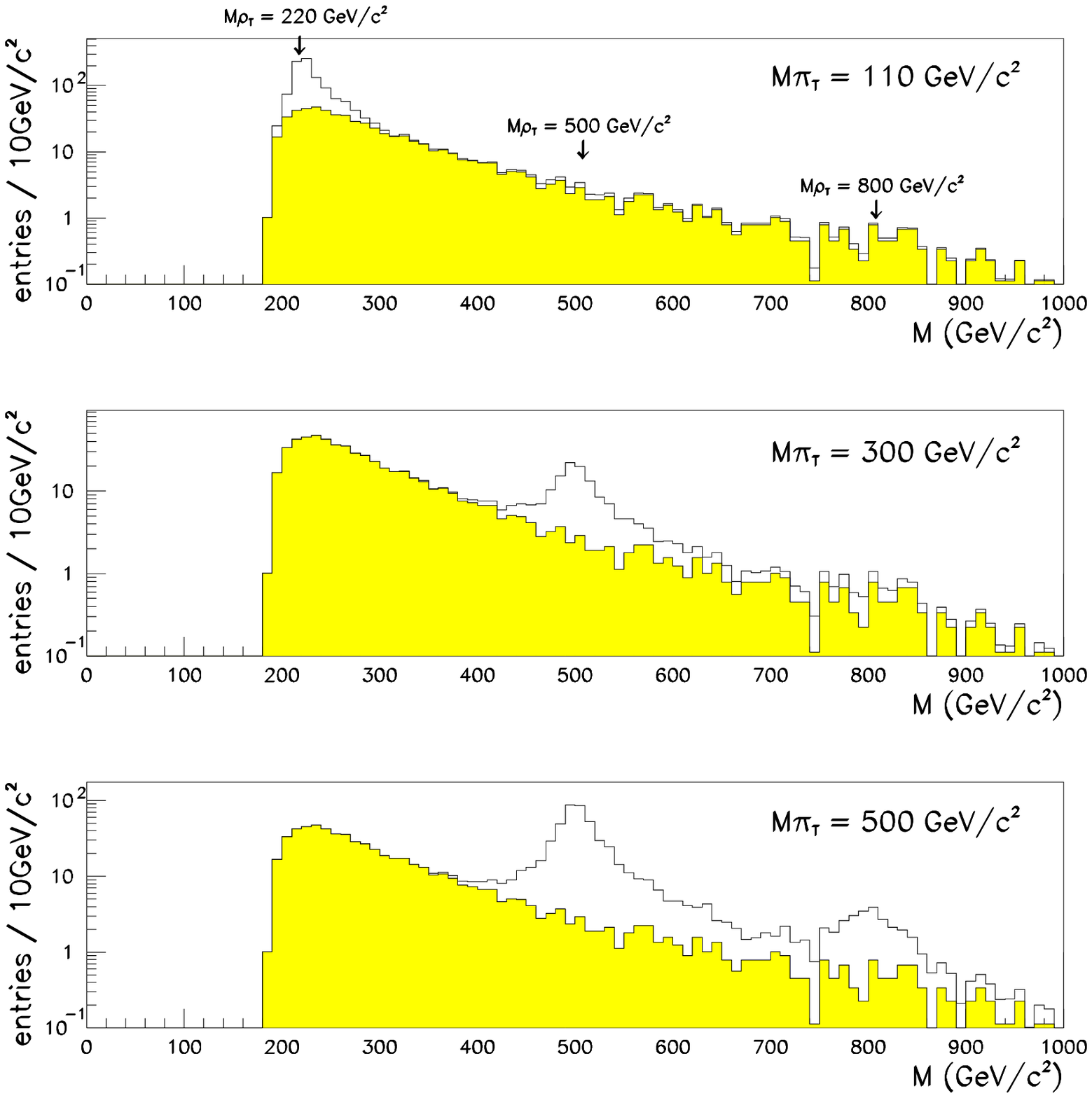}
\vskip5.0truecm
 \caption{\it
      Simulated event and background rates in the ATLAS detector for $\tropm
      \ra W^\pm Z \ra \ell^\pm\nu_\ell \ell^+\ell^-$ for various $M_{\tro}$
      and $M_{\tpi}$; from Ref.~\cite{atlastdr}.
    \label{fig16} }
\end{figure}
\section{Technicolor Phenomenology}

The coupling $\atc$ in walking technicolor decreases slowly if the
beta--function $\beta(\atc) = \mu \ts d\atc/d\mu$ is negative and near zero
for a large range of energy $\mu$ above $\LTC$. This small $\beta$--function
may be achieved by having many technifermions in the fundamental
representation of $\sutc$, or a few in higher-dimensional representations, or
both~\cite{multiklee}. For different reasons, models of topcolor--assisted
technicolor also seem to require many
technifermions~\cite{tctwoklee,tctwokl}. The technidoublets include $\sim$ 5
that are color singlets as well as the color triplets $T_1 \in (\Ntc,3,1)$
and $T_2 \in (\Ntc,1,3)$ mentioned above. The color singlets insure
that all quarks and leptons get the appropriate ETC mass and that there is
sufficient mixing between the third generation quarks and the two light ones
(so that weak decays of the $b$--quark are allowed).

These requirements suggest that the technicolor scale is much lower than
previously thought. If the number $N$ of technidoublets is $\CO(10)$
(including 3~for each color triplet), then $\LTC \simeq F_T = F_\pi/\sqrt{N}
\simle 100\,\gev$. This sets the mass scale for the lightest {\it
color--singlet} technivector mesons, $M_{\tro} \simeq M_{\tom} \simeq 2\LTC
\simle 200\,\gev$. These states are produced in hadron and lepton
colliders. The mechanism is good old--fashioned vector meson dominance of the
$s$--channel production of $\gamma$, $Z^0$, and $W^\pm$. The lightest
color--octet $\troct$, bound states of the color--triplet technifermions of
TC2, will be heavier, starting, perhaps, at 400--500~GeV. Hadron colliders
are needed to produce these states. Isosinglet $\troct$ bosons are produced
by their couplings to the QCD gluon and the $V_8$ colorons.

In the limit that color interactions are weak compared to technicolor, the
chiral symmetry of $N$ technidoublets is $SU(2N)_L \otimes SU(2N)_R$. When it
is spontaneously broken, there result $4N^2 - 4$ technipions in addition to
$W^\pm_L$ and $Z^0_L$, a large number of states if $N$ is large. In QCD--like
technicolor, these technipions would be very light and the $\tro$ and $\tom$
would decay to two or more technipions, with $\troct$ decaying to
color--octet and color--triplet (leptoquark) pairs. Walking technicolor
dramatically changes this expectation. In the extreme walking limit,
$\condetc \simeq (\METC/\LTC) \condtc$, so that technipions masses are of
order $\LTC$, and they are not pseudoGoldstone bosons at all. Though this
extreme limit is theoretically problematic because it is exactly
scale--invariant, it is clear that walking technicolor enhances $\tpi$ masses
significantly more than it does the $\tro$ and $\tom$ masses. Thus, it is
likely that $M_{\tpi} \simge \half M_{\tro,\tom}$ and, so, the nominal
isospin--conserving decay channels $\tro \ra \tpi\tpi$ and $\tom \ra
\tpi\tpi\tpi$ are {\it closed}~\cite{multiklee}. If the color--singlet $\tro$
start near $200\,\gev$, we expect $M_{\tpi} \simge 100\,\gev$. Of course, an
explicit ETC model will be needed to make firm mass estimates.

This ``low--scale technicolor'' may be within the reach of CDF and D\O\ in
Run~II of the Tevatron
Collider~\cite{elw,tcsm_singlet,tcsm_octet}.~\footnote{Many of these
signatures are now encoded in {\sc Pythia}~\cite{pythia}.} It certainly will
be accessible at the LHC. Color--singlet $\tro$ and $\tom$ may even be
detected at LEP200. If a lepton collider with $\ecm \simle 500\,\gev$ is
built, it will be able carry out precision studies of color--singlet
technihadrons. The Very Large Hadron Collider or a multi--TeV lepton collider
will be needed to explore more fully the strongly coupled region of walking
technicolor.

In the rest of this section, we describe a simple model, suitable for
experimental studies, of our expectations for the low--lying states of
low--scale technicolor---first for the the color--singlet sector, then for
color--nonsinglets.

\subsection{\it {\underbar{Theory and Experiment for Color--Singlet
        Technihadrons}}}

The flavor problem is hard whether it is attacked with extended technicolor
or from any other direction. We theorists need experimental guidance.
Experimentalists, in turn, need input from theorists to help design useful
searches.  Supersymmetry has its MSSM. What follows is a description of the
corresponding thing for technicolor, in the sense that it defines a set of
incisive experimental tests in terms of a limited number of adjustable
parameters. I call this the ``Technicolor Straw Man'' model (or TCSM).

In the TCSM, we assume that we can consider {\it in isolation} the
lowest-lying bound states of the lightest technifermion doublet, $(T_U,
T_D)$. If these technifermions belong to the fundamental representation of
$\sutc$, they probably are color singlets. In walking technicolor, ordinary
color interactions contribute significantly to the hard mass of $SU(3)$
triplets~\cite{multiklrm}. The lightest technidoublet's electric charges are
unknown; we denote them by $Q_U$ and $Q_D = Q_U-1$.

The bound states in question are vector and pseudoscalar mesons. The vectors
include a spin--one triplet $\tro^{\pm,0}$ and a singlet $\tom$. In
topcolor--assisted technicolor, there is no need to invoke large
isospin--violating extended technicolor interactions to explain the
top--bottom splitting. Techni--isospin can be, and likely must be, a good
approximate symmetry. Then, $\tro$ and $\tom$ will be mostly isovector and
isoscalar, respectively, and they will be nearly degenerate. Their production
in $\ol q q$ and $e^+e^-$ annihilation is described using vector meson
dominance, with propagator matrices that mix them with $W^\pm$ and $\gamma$,
$Z^0$. The details are given in Ref.~\cite{tcsm_singlet}, called TCSM--1
below. I reiterate, mixing of these $\tro$ and $\tom$ with their excitations
is ignored in the TCSM as is the production of the axial vector $a_{1T}$ and
the like.

The lightest pseudoscalar $\ol T T$ bound states, the technipions, also
comprise an isotriplet $\Pi_T^{\pm,0}$ and an isosinglet $\Pi_T^{0
\prime}$. However, these are not mass eigenstates; all color--singlet
isovector technipions have a $W_L$ component. To limit the number of
parameters in the TCSM, we make the simplifying assumption that the
isotriplets are simple two--state mixtures of the $W_L^\pm$, $Z_L^0$ and the
lightest mass eigenstate pseudo--Goldstone technipions $\tpi^\pm, \tpiz$:
\be\label{eq:pistates}
 \vert\Pi_T\rangle = \sin\chi \ts \vert
W_L\rangle + \cos\chi \ts \vert\tpi\rangle\ts.
\ee
Here, $\sin\chi = F_T/F_\pi = 1/\sqrt{N} \ll 1$ is an adjustable
parameter. The isosinglet is also an admixture, $\vert\Pi_T^{0 \prime}
\rangle = \cos\chipr \ts \vert\tpipr\rangle\ + \cdots$, where $\chipr$ is
another adjustable mixing angle and the ellipsis refers to other technipions
needed to eliminate the two-technigluon anomaly from the $\Pi_T^{0 \prime}$
chiral current.

It is unclear whether, like $\troz$ and $\tom$, the neutral technipions
$\tpiz$ and $\tpipr$ will be degenerate as we have previously
supposed~\cite{elw}. On one hand, they both contain the lightest $\ol T T$ as
constituents. On the other, $\tpipr$ must contain other, presumably heavier,
technifermions as a consequence of anomaly cancellation. The calculations and
searches presented here assume that $\tpiz$ and $\tpipr$ are nearly
degenerate. If this is true, and if their widths are roughly equal, there
will be appreciable $\tpiz$--$\tpipr$ mixing. Then, the lightest neutral
technipions will be ideally-mixed $\ol T_U T_U$ and $\ol T_D T_D$ bound
states.

In any case, these technipions are expected to couple most strongly to the
heaviest fermion pairs that they can. The reason for this is that $\tpi$
couple to ordinary fermions via extended technicolor, $\tpi \ra \ol T T \ra
\ol f f$. Figure~1 suggests that this coupling is proportional to $m_f$ (more
precisely, the ETC contribution to $m_f$). In our studies we assume
technipions to be lighter than $m_t + m_b$. Then, we expect them to decay as
follows: $\tpip \ra c \ol b$ or $c \ol s$ or even $\tau^+ \nu_\tau$; $\tpiz
\ra b \ol b$ and, perhaps $c \ol c$, $\tau^+\tau^-$; and $\tpipr \ra gg$, $b
\ol b$, $c \ol c$, $\tau^+\tau^-$.~\footnote{See Ref.~\cite{tcsm_singlet} for
a discussion and estimate of $\tpi$ decay rates.} This puts a premium on
heavy--flavor identification in collider experiments. However, this is only
an educated guess. The reader is cautioned that the mass--eigenstate neutral
$\tpi$ may have a sizable branching ratio to gluon (or even light--quark)
pairs.


For vanishing electroweak couplings $g$ and $g'$, the $\tro$ and $\tom$ decay
as
\bea\label{eq:vt_decays}
\tro &\ra& \Pi_T \Pi_T = \cos^2 \chi\ts (\tpi\tpi) + 2\sin\chi\ts\cos\chi
\ts (W_L\tpi) + \sin^2 \chi \ts (W_L W_L) \ts; \nn \\
\tom &\ra& \Pi_T \Pi_T \Pi_T = \cos^3 \chi \ts (\tpi\tpi\tpi) + \cdots \ts.
\eea
As noted above however, the all--$\tpi$ modes are likely to be closed. Thus,
major decay modes of the $\tro$ will be $W_L\tpi$ or, if $M_{\tro} \simle
180\,\gev$ (a possibility we regard as unlikely, if not already eliminated by
LEP data), $W_L W_L$. The $W^\pm \tpi^{\mp, 0}$ and $Z^0 \tpi^\pm$ decays
of $\tro$ have striking signatures in any collider. Only at LEP is it now
possible to detect $\troz \ra W^+W^-$ above the standard model background. If
$M_{\tom} < 250\,\gev$, all the $\tom \ra \Pi_T \Pi_T \Pi_T$ modes are
closed. In all cases, the $\tro$ and $\tom$ are very narrow, $\Gamma(\tom)
\simle \Gamma(\tro) \simle 1\,\gev$, because of the smallness of $\sin\chi$
and the limited phase space. Therefore, we must consider other decay
modes. These are electroweak, suppressed by powers of $\alpha$, but not by
phase space.

The decays $\tro, \tom \ra G \tpi$, where $G$ is a {\it transversely}
polarized electroweak gauge boson, and $\tro, \tom \ra \ol f f$ were
calculated in TCSM--1. The $G \tpi$ modes have rates of $\CO(\alpha)$, while
the fermion mode $\ol f f$ rates are $\CO(\alpha^2)$. The $\Gamma(\tro, \tom
\ra G \tpi)$ are suppressed by $1/M^2_V$ or $1/M^2_A$, depending on whether
the vector or axial vector part of the electroweak current is involved in the
decay. Here, $M_{V,A}$ are masses of order $\LTC$ occuring in the
dimension--5 operators for these decays. We usually take them equal and vary
them from 100 to 400~GeV. For the smaller values of $M_{V,A}$, these modes,
especially $\tro, \tom \ra \gamma\tpi$, are as important as the $W_L\tpi$
modes. For larger $M_{V,A}$ and $|Q_U + Q_D| \simge 1$, the $\ol f f$ decay
modes may become competitive. As an illustration, Table~1 lists the relative
strengths of the decay amplitudes for the $\tro, \tom \ra G \tpi$
processes. Figure~2 gives a sense of the $M_{V,A}$ dependence of the total
decay rates of $\tro$ and $\tom$ for $M_{\tro} = 210\,\gev$, $M_{\tom} =
200$--$220\,\gev$, $M_{\tpi} = 110\,\gev$, and $Q_U = Q_D + 1 =
4/3$. Figure~3 shows the decay rates for $Q_U = -Q_D = 1/2$. Note how narrow
the $\tro$ and $\tom$ are. These and all subsequent calculations assume that
$\Ntc = 4$ and $\sin\chi=\sin\chi'=1/3$.  Experimental analyses quoted below
use the same defaults and (usually) $Q_U = Q_D + 1 = 4/3$.

Figures~4 and~5 show the cross sections in $\ol p p$ collisions at
$\ecm=2\,\tev$ for production of $\gamma\tpi$ and for $W\tpi$, $Z\tpi$ and
$\tpi\tpi$ as a function of $M_V = M_A$. Figure~6 shows the $e^+e^-$ rate for
$M_V = 100\,\gev$. The production rates in these figures, all in the picobarn
range, are typical for the Tevatron for $M_{\tro,\tom} \simle 250\,\gev$ and
$M_{\tpi} \simle 150\,\gev$. That is why we believe Run~II will probe a
significant portion of the parameter space of low--scale technicolor.

Let us turn to the recent searches for color--singlet technihadrons. We begin
with analyses by the L3~\cite{L3} and DELPHI~\cite{delphi} collaborations at
LEP. Note that the LEP experiments can be sensitive to $\tro$ and $\tom$
masses significantly above the $e^+e^-$ c.m.~energy, $\ecm$. This is because
the $e^+e^-$ cross section on resonance is very large for the narrow
$\tro$. Furthermore, masses below the nominal c.m.~energy are scanned by the
process of radiative return, $e^+e^- \ra \troz/\tom + n\ts\gamma$.

The L3 search is based on $176\,\ipb$ of data taken at an average energy of
$189\,\gev$. The analysis used TCSM--1 for the channels $e^+e^- \ra \troz \ra
W^+W^-$; $W^\pm_L \tpimp \ra \ell\nu_\ell bc$; $\tpip\tpim \ra c\ol b \ts b
\ol c$; and $\gamma \tpiz \ra \gamma b \ol b$. The TC--scale masses were
fixed at $M_V = M_A = 200\,\gev$ and the technifermion charges ranged over
$Q_U + Q_D = 5/3,0,-1$. The resulting 95\% confidence limits in the
$M_{\tro}$--$M_{\tpi}$ plane are shown in Fig.~7.

The DELPHI collaboration searched for $\troz \ra W^+W^-$; $W^\pm \tpimp$;
$\tpip\tpim$; and $\mu^+\mu^-$. Data was taken over a range of $\ecm$ between
161 and $202\,\gev$ with a variety of integrated luminosities. The modes
$\troz,\tom \ra \tpiz \gamma$, $\tpipr\gamma$ were neglected. This is not a
good assumption if $M_V \simle 200\,\gev$. The DELPHI exclusion plot is shown
in Fig.~8.

Since these LEP analyses were done, I have realized that the cross section
formulae stated in TCSM--1 are inappropriate for $\ecm$ well below
$M_{\tro}$.  This is unimportant for the Tevatron and LHC, where the
production rate comes mainly from integrating parton distributions over the
resonance pole. However, it may have a significant effect on limits derived
from $e^+e^-$ annihilation. This is especially true for the $W^+W^-$ channel,
which has a large standard model amplitude interfering with the TCSM
one.\footnote{I thank F.~Richard for drawing my attention to this shortcoming
of TCSM--1. A correction will be issued soon.} Another feature of these
analyses not evident in the exclusion plots is that limits on $M_{\tpi}$
approaching $\ecm/2$ should be derivable from $e^+ e^- \ra \tpi^+\tpi^-$. We
look forward to the new LEP limits that will be announced in the summer of
2000.

Both Tevatron collider collaborations have searched for signals of low--scale
technicolor. In Run~I, only CDF had a vertex detector to find the detached
vertices of $b$--quark decays. The collaboration used this capability to
search for processes signalled by a $W$ or photon plus two jets, one of which
is $b$--tagged:
\bea\label{eq:cdfsearches}
\ol q q \ra W^\pm,\gamma,Z^0 &\ra& \tro^{\pm,0} \ra W^\pm_L \tpi \ra \ell^\pm
\nu_\ell \ts b + \jet \nn\\
        &\ra& \tro^{\pm,0}, \tom \ra \gamma \tpi \ra
        \gamma \ts b + \jet \ts.
\eea
These analyses were carried out before the publication of TCSM--1, so they do
not include the $G\tpi$ and $\ol ff$ processes and corresponding branching
ratios. They will be included in Run~II data analyses. Figure~9 shows data
for the $W\tpi$ search on top of a background and signal expected for default
parameters with $M_{\tro} = 180\,\gev$ and $M_{\tpi} = 90\,\gev$. The
topological cuts leading to the lower figure are described in the second
paper of Ref.~\cite{elw}. The region excluded at 95\% confidence level is
shown in Fig.~10~\cite{cdfwpi}.

Figure~11 shows the invariant mass of the tagged and untagged jets and the
invariant mass difference $M(\gamma+b+\jet) - M(b+\jet)$ in a search for
$\tom,\tro \ra \gamma \tpi$~\cite{cdfgpi}. The good resolution in this mass
difference is controlled mainly by that of the electromagnetic energy.  The
exclusion plot is shown in Fig.~12. It is amusing that the $\sim 2\sigma$
excesses in Figs.~9 and 12 are both consistent with expectations for a signal
with $M_{\tro,\tom}\simeq 200\,\gev$ and $M_{\tpi}\simeq 100\,\gev$.

The expected reach of CDF in Run~IIa for the $\tro \ra W^\pm \tpi \ra
\ell^\pm \nu_\ell \ts b \ts {\rm jet}$ processes is shown in Fig.~13 for $M_V
= M_A = 200\,\gev$~\cite{handa}. This study uses all the processes of
TCSM--1. It also assumes the same selections and systematic uncertainty as in
the published Run~I data~\cite{cdfwpi}, but double the signal efficiency
(1.38\% vs. 0.69\%). The $5\sigma$ discovery reach goes up to $M_{\tro} =
210\,\gev$ and $M_{\tpi} = 110\,\gev$, larger than the 95\% excluded region
in Run~I. The region that can be excluded in Run~IIa extends up to $M_{\tro}
= 250\,\gev$ and $M_{\tpi} = 145\,\gev$. When $M_V = 400\,\gev$, the
$5\sigma$ discovery and 95\% exclusion regions are only slightly larger than
this.

The Run~I D\O\ detector had superior calorimetry and hermiticity. The
collaboration studied its Drell--Yan data to search for $\tro, \tom \ra
e^+e^-$~\cite{dzeroee}. The data and the excluded region are shown in Fig.~14
for $Q_U = Q_D + 1 = 4/3$, $M_V = 100$--$400\,\gev$ and $M_{\tro} - M_{\tpi}
= 100\,\gev$. Increasing $M_V$ and decreasing $M_{\tro} - M_{\tpi}$ both
increase the branching ratio for the $e^+e^-$ channel. For the parameters
considered here, $M_{\tro} = M_{\tom} < 150$--$200\,\gev$ is excluded at the
95\% CL. The expected reach of D\O\ in Run~IIa for $\tro,\tom \ra e^+e^-$
with $M_V = 100$ and $200\,\gev$ and other TCSM parameters (see above) is
shown in Fig.~15~\cite{meena}. As long as $Q_U + Q_D = \CO(1)$, masses
$M_{\tro,\tom}$ up to 450--500~GeV should be accessible in the $e^+e^-$
channel.

The ATLAS collaboration has studied its reach for $\tro \ra W^\pm Z, \ts
W^\pm \tpi, \ts Z \tpi$ and for $\tom \ra \gamma\tpi$~\cite{atlastdr}.
Figure~16 shows $\tropm \ra W^\pm Z \ra \ell^\pm \nu_\ell \ell^+\ell^-$ for
several $\tro$ and $\tpi$ masses and a luminosity of $10\,\ifb$. Detailed
studies have not been published in which all the TCSM processes have been
included and the parameters varied over a wide range. Still, it is clear from
Fig.~16 that the higher energy and luminosity of the LHC ought to make it
possible to completely exclude, or discover, low--scale technicolor for any
reasonable parameterization.

\subsection{\it {\underbar{Color--Nonsinglet Technihadrons}}}
\begin{figure}[t]
\vspace{9.0cm}
\includegraphics{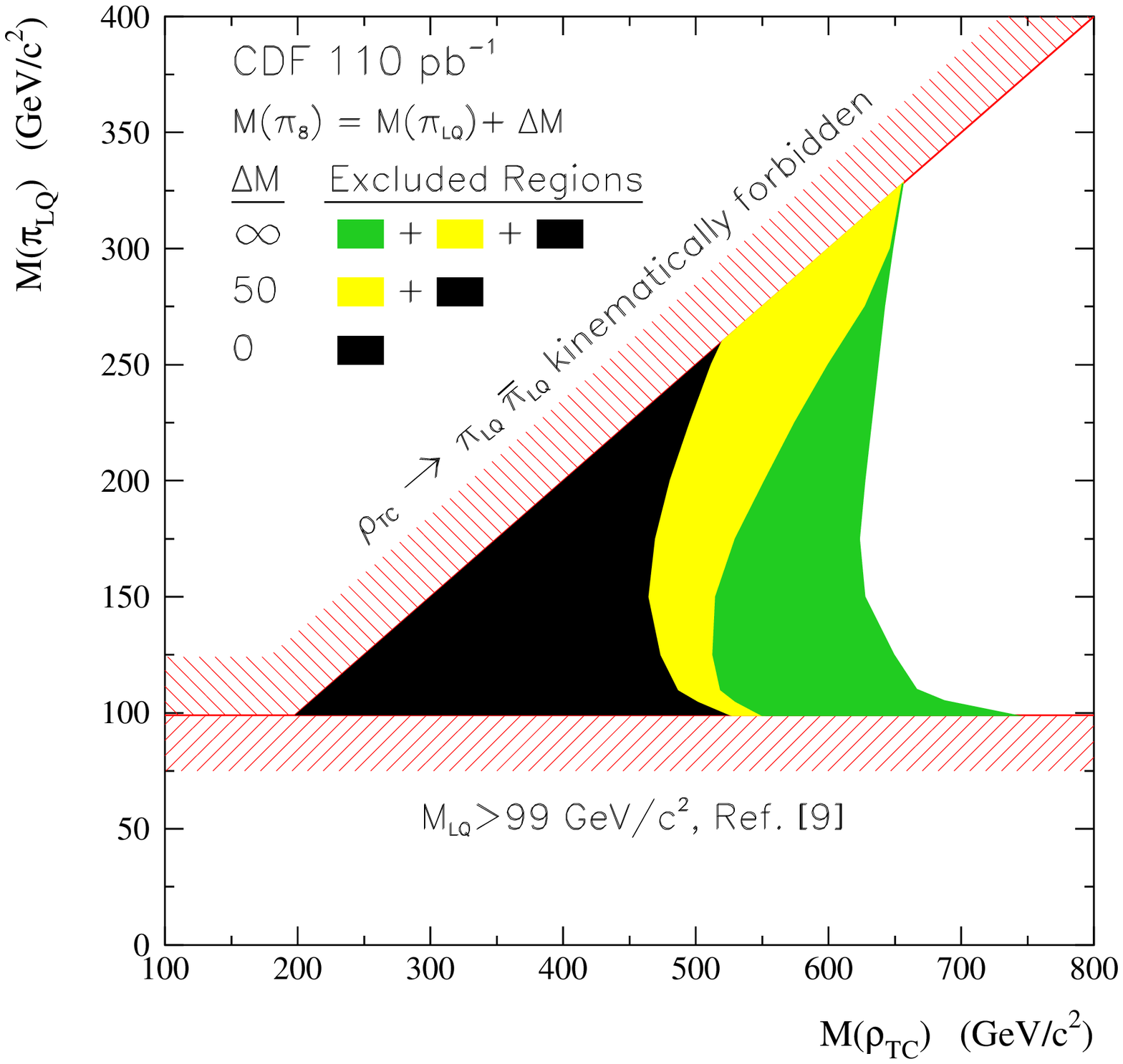}
\vskip2.8truecm
\caption{\it
  The 95\% CL exclusion regions for various $M_{\octpi} - M_{\tpilq}$ from a
  CDF search for $\troct \ra \tpilq\tpiql \ra \tau^+\tau^-\ts\jet\ts\jet$;
  from Ref.~\cite{cdfed}.
\label{fig17} }
\end{figure}
\begin{figure}[t]
 \vspace{9.0cm}
\includegraphics{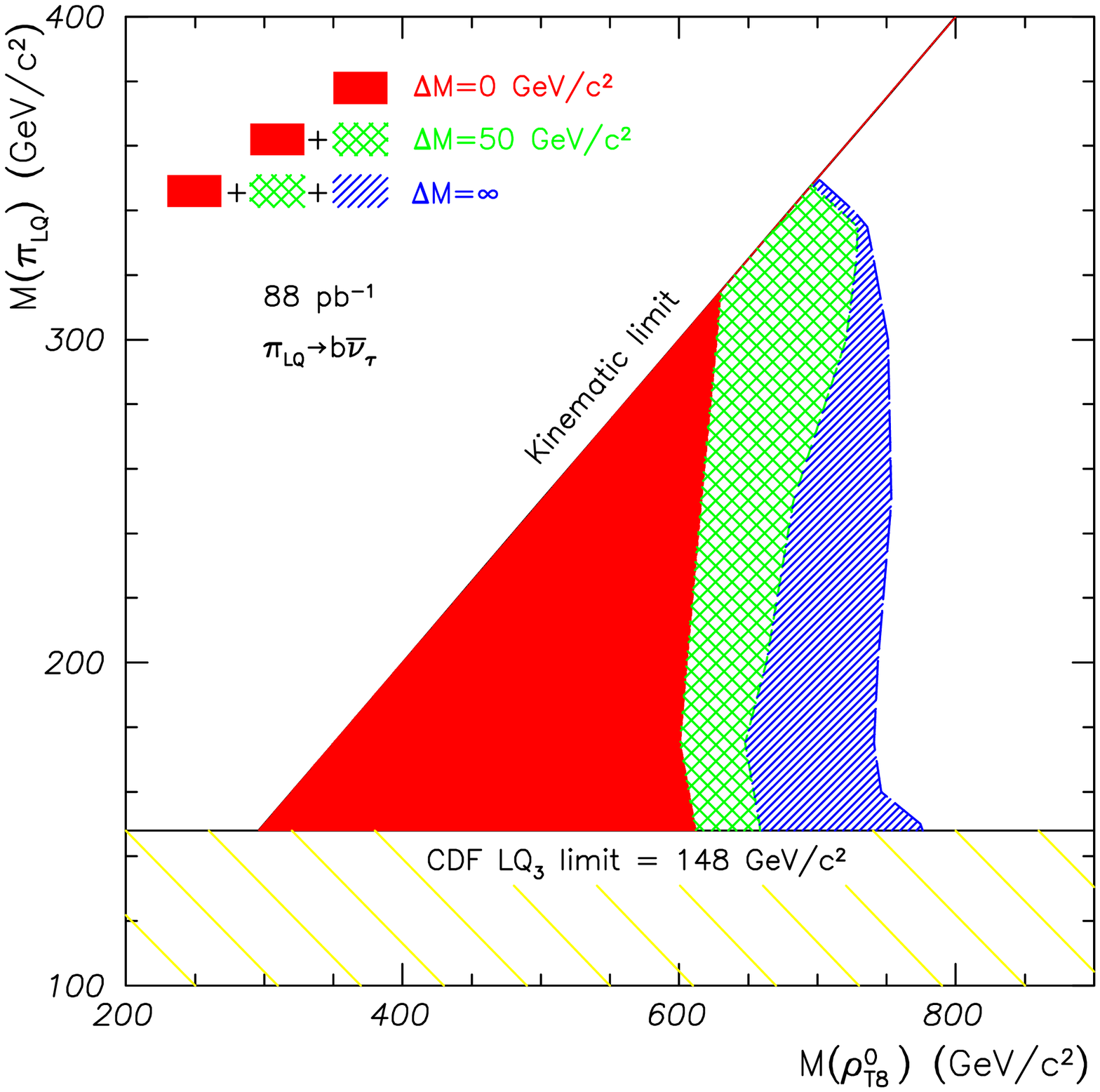}
\vskip2.7truecm
 \caption{\it
   The 95\% CL exclusion regions for various $M_{\octpi} - M_{\tpilq}$ from a
  CDF search for $\troct \ra \tpilq\tpiql \ra \ol b b \nu\nu$; from
  Ref.~\cite{cdfnd}.
    \label{fig18} }
\end{figure}
\begin{figure}[t]
 \vspace{9.0cm}
\includegraphics{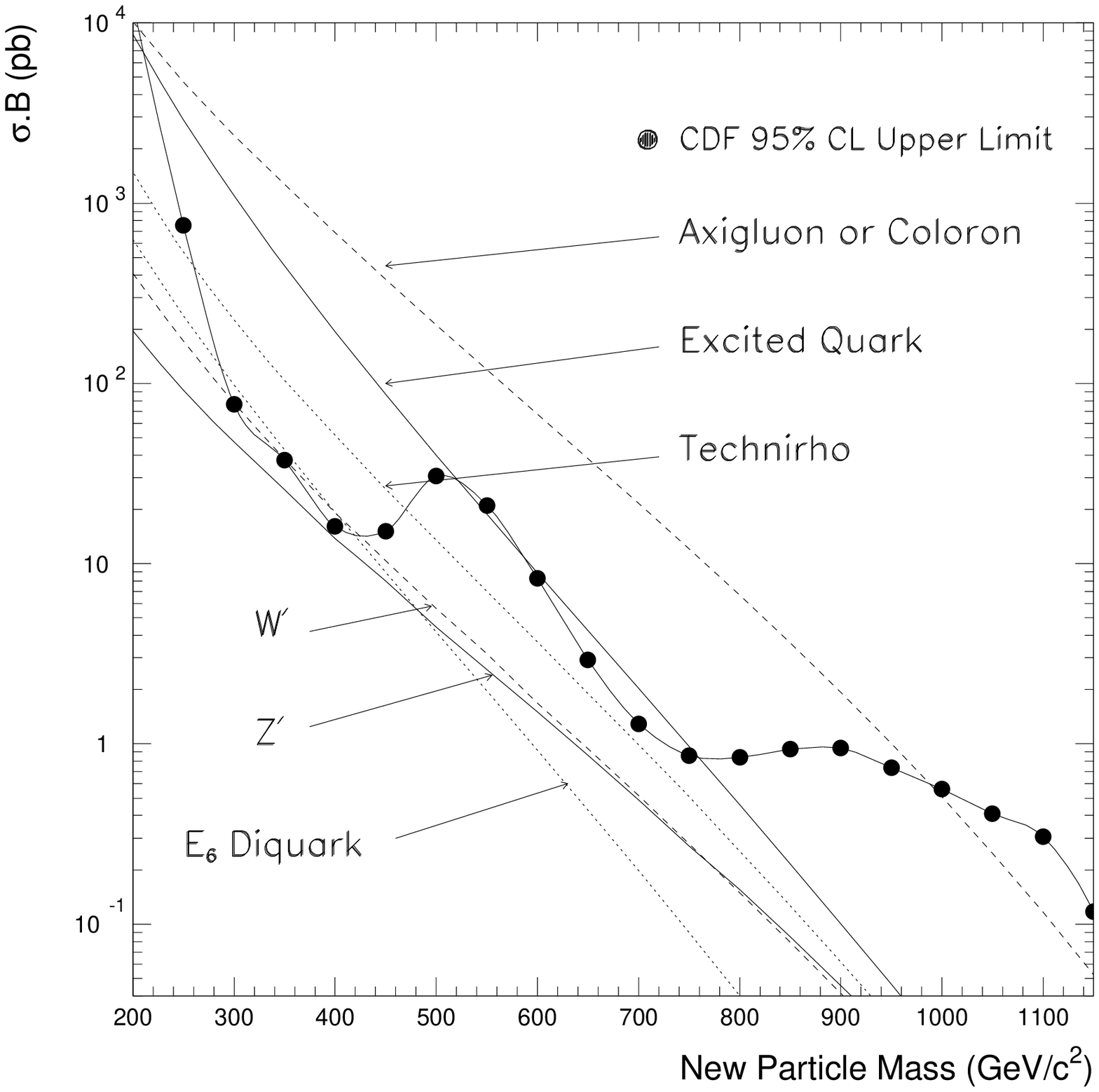}
\vskip1.8truecm
 \caption{\it
      The 95\% exclusions for a CDF search for $\troct \ra \jet\ts\ts\jet$ and
      other narrow dijet resonances; from Ref.~\cite{cdfjets}.
    \label{fig19} }
\end{figure}
\begin{figure}[t]
 \vspace{9.0cm}
\includegraphics{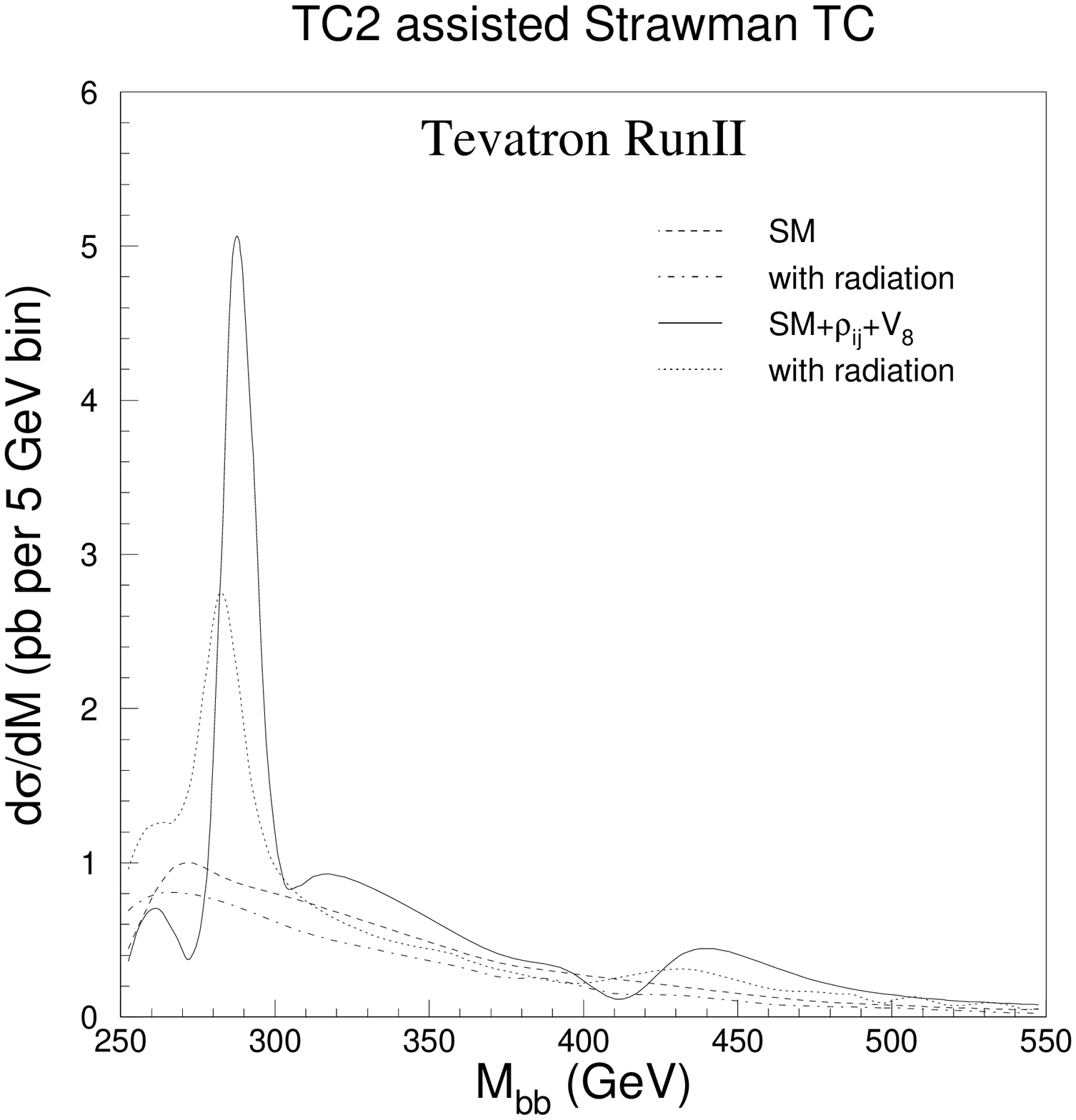}
\vskip1.8truecm
 \caption{\it
      Production of $\ol bb$ in $\ol pp$ collisions at $\ecm=2\,\tev$
      according to the TCSM model of Ref.~\cite{tcsm_octet}.
    \label{fig20} }
\end{figure}

\begin{figure}[t]
 \vspace{9.0cm}
\includegraphics{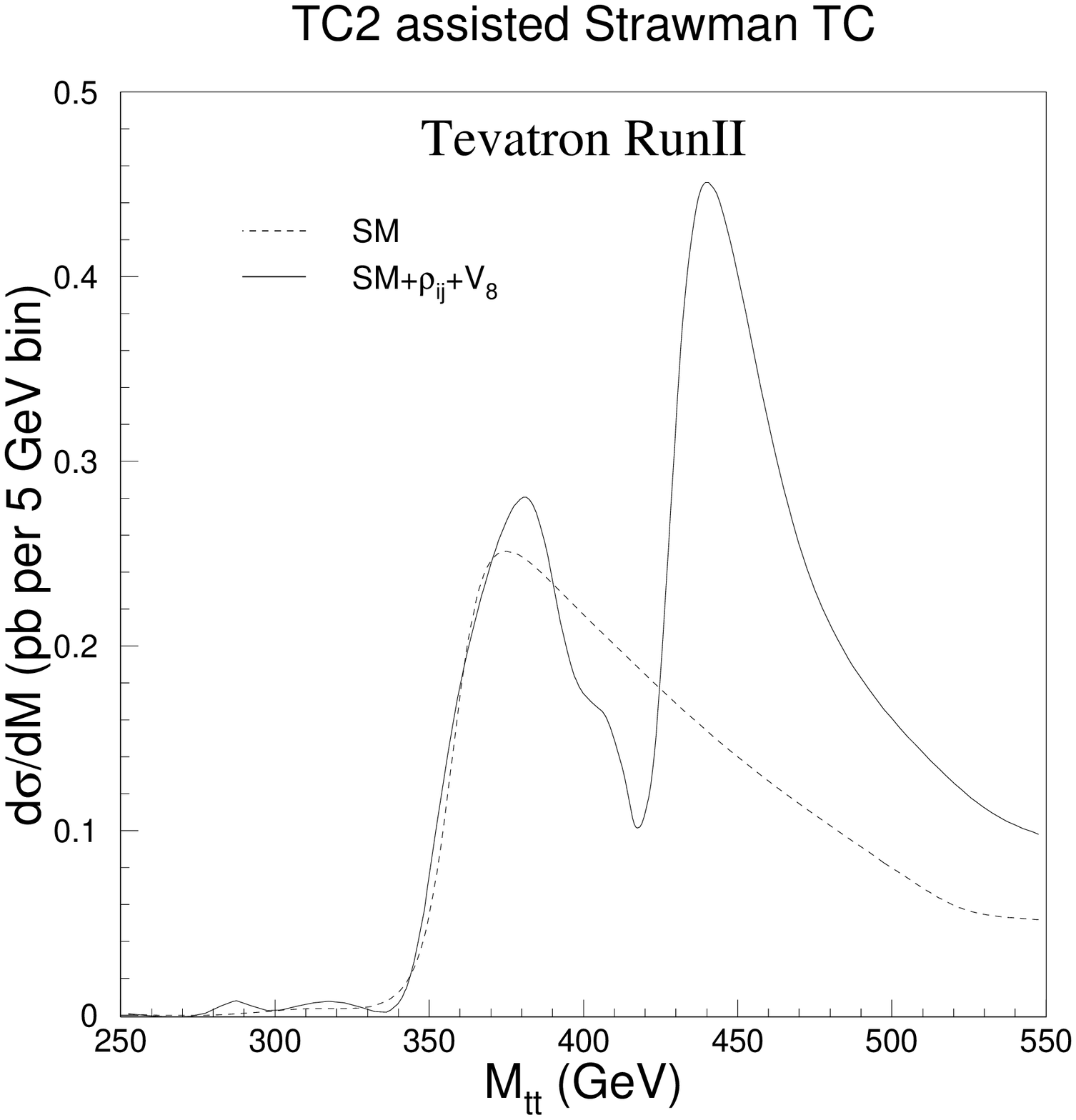}
\vskip1.8truecm
 \caption{\it
      Production of $\ol tt$ in $\ol pp$ collisions at $\ecm=2\,\tev$
      according to the TCSM model of Ref.~\cite{tcsm_octet}.
    \label{fig21} }
\end{figure}

The experimental searches so far in the color--nonsinglet sector of
low--scale technicolor have been inspired by the phenomenology of a
pre--TCSM, one--family TC model. This model contains a single doublet each of
color--triplet techniquarks $Q = (U,D)$ and of color--singlet technileptons
$L = (N,E)$~\cite{fs,ehlq,multiklrm}. We consider these searches first,
commenting on the status of the TCSM for color nonsinglets at the end.

Assuming that techni--isospin is conserved, production of color--nonsinglet
states is assumed to proceed through the lightest isoscalar color--octet
technirho, $\troct$:
\bea\label{eq:nonsing}
\ol q q, \ts gg  \ra g \ra \troct &\ra& \octpi\octpi,\ts\ts \tpilq\tpiql
\nn\\
        &\ra& \ol q q, \ts gg \ts\ts\ts {\rm dijets} \ts.
\eea
Here, $\octpi = \octpipm, \octpiz, \pi_{T8}^{0'}\equiv\eta_T$ are four
color--octet technipions that are expected to decay to heavy $\ol q q$ pairs;
$\tpilq$ are four color--triplet leptoquarks expected to decay to heavy
$\ol \ell q$ with the corresponding charges. If TC2 is invoked, the neutral
$\octpi$ decay to $\ol b b$ and, possibly, $gg$ as readily as to $\ol t
t$.~\footnote{The ATLAS collaboration has studied $gg \ra \eta_T \ra \ol t
t$~\cite{atlastdr}. Even if this is not the dominant decay mode of an
$\eta_T$ in a TC2 model, other bosons, such as the colorons $V_8$, will have
a sizable $\ol tt$ branching ratio and the ATLAS study serves as a promising
prototype of a search for this process.}

The two $\troct \ra \tpi\tpi$ searches are by CDF for leptoquark technipions:
$\tpied \ra \tau^+ b$ where the $b$ is not tagged~\cite{cdfed} and $\tpind
\ra \nu b$, $\nu c$ with one or two tagged jets~\cite{cdfnd}. These are based
on $110\,\ipb$ and $88\,\ipb$ of Run~I data, respectively. The exclusion plot
for the $\tau^+\tau^- +$dijet signal is shown in Fig.~17 as a function of the
$\octpi$--$\tpilq$ mass difference. The theoretically likely case is that
this mass difference is about 50~GeV, implying a 95\% excluded region
extending over $200 \simle M_{\troct} \simeq 2 M_{\tpilq} \simle
500\,\gev$. Figure~18 shows the reach for $\tros \ra b \ol b \nu \ol\nu$ with
at least one $b$--jet tagged. Here the 95\% limits extends over $300 \simle
M_{\troct} \simeq 2 M_{\tpilq} \simle 600\,\gev$. The search for $\tpilq \ra
c \nu$ excludes a similar range. These limits are quite impressive. However,
it is not clear how they will be affected by the complications of
topcolor--assisted technicolor.

Given the walking technicolor enhancement of $\tpi$ masses, it is likely that
the $\troct \ra \tpi\tpi$ channels are closed. In that case, one looks for
$\troct$ as a dijet resonance: $\ol qq, \ts gg \ra g \ra \troct \ra g \ra \ol
qq, gg$.~\footnote{The decays $\troct \ra g \octpi$ and $g \pi_{T1}$ may
occur and deplete the $\troct \ra {\rm dijets}$ rate. These modes were not
expected to be important in Ref.~\cite{multiklrm} and they were not taken
into account in the CDF analyses. They are included in the studies in
Ref.~\cite{tcsm_octet} (TCSM--2) discussed below.}  Searches have been made
by CDF for both untagged~\cite{cdfjets} and $b$--tagged~\cite{cdfbbjets}
dijet resonances. The latter mode has a better signal--to--background ratio,
but the rates and $b$--identification efficiencies in Run~I were not high
enough to make this advantage significant; the best limits come from
untagged--dijets. The results of such a search are shown in Fig.~19. The
region $260 < M_{\troct} < 460\,\gev$ is excluded at the 95\% confidence
level. This is a stringent constraint, but its applicability to TC2 models is
uncertain.

Finally, and very briefly, we turn to the effect of topcolor--assisted
technicolor on experimental studies of the color--nonsinglet
sector~\cite{tcsm_octet}. As I mentioned, the simplest implementation of TC2
models requires two color $SU(3)$ groups, one strongly--coupled at 1~TeV for
the third generation quarks $(t,b)$ and one weakly--coupled for the two light
generations. These two color groups must be broken down to the diagonal
$SU(3)$ near 1~TeV, and this remaining symmetry is identified with ordinary
color. The most economical way I know to achieve this is to have the two
technifermion doublets $T_1= (U_1,D_1) \in (\Ntc,3,1)$ and $T_2 = (U_2,D_2)
\in (\Ntc,1,3)$ condense with each other to achieve the desired breaking to
$SU(3)_C$~\cite{tctwokl}.

The main phenomenological consequence of this scenario is that the $\suc$
gluons mix with the $\suone$ octet of massive colorons, $V_8$, and with four
color--octet technirhos, $\rho_{ij} \sim \ol T_i \lambda_A T_j$ ($i, j =
1,2$)~\cite{tcsm_octet}. The colorons decay strongly to top and bottom quarks
and weakly to the light quarks~\cite{tctwohill}. Alternatively, in the
flavor--universal variant of TC2~\cite{ccs}, the colorons decay with equal
strength to all quark flavors. In Ref.~\cite{tcsm_octet}, we assume for
simplicity that all $\rho_{ij}$ are too light to decay to pairs of
technipions.\footnote{The colored technipion sector of a TC2 model is bound
to be very rich. Thus, it is not clear how the limits on leptoquarks
discussed above are to be interpreted. This is work for the future.} Then,
they decay (via gluon and coloron dominance) into $\ol q q$ and $gg$ dijets
and into $g \octpi$ and $g\pi_{T1}$.

Even this simplified minimal TC2 version of the TCSM has a much richer set of
dijet spectra and other hadron collider signals than the one--family model
discussed above~\cite{multiklrm,cdfjets}. We are just beginning to study
it. Some preliminary examples of dijet production based on the assumptions of
TCSM--2 are shown in Figs.~20 and~21 for $\ecm = 2\,\tev$ at the Tevatron. In
both figures the coloron mass is 1.2~TeV while the input $\troct$ masses
range from 350 to 500~GeV.~\footnote{The pole masses are shifted somewhat
from these input values by mixing effects.} Figure~20 shows $\ol b b$
production with a strong resonance at 300~GeV (i,e., below $\ol t t$
threshold). Figure~21 shows $\ol t t$ production with roughly a factor of
two enhancement of the total cross section over that predicted in the
standard model. Both signals are ruled out by Run~I measurements of the $\ol
b b$ and $\ol t t$ cross sections~\cite{cdfbbjets,ttlimit}.

Many more studies of both the color--singlet and nonsinglet sectors of the
TCSM need to be carried out. The Fermilab Workshop on Strong Dynamics at
Run~II will begin these studies this autumn, in time to be of use when the
run starts in Spring~2001. The CDF and D\O\ collaborations will carry out
detector--specific simulations in the next year or two. More detailed and
more incisive $e^+e^- \ra \tro,\tom$ studies will come from the LEP
experiments this year. The ATLAS and CMS collaborations likewise ought to
study a broad range of signals for strong dynamics before they begin their
runs later in the decade.

\section{Open Problems}

My main goal in these lectures has been to attract some bright young people
to the dynamical approach to electroweak and flavor symmetry breaking. Many
difficult problems remain open for study there. These lectures provide a
basis for starting to tackle them. All that's needed now are new ideas, new
data, and good luck. Here are the problems that vex me:

\begin{enumerate}

\item{} First and foremost, we need a reasonably realistic model of extended
technicolor, or {\it any other} natural, dynamical description of flavor. To
repeat: This is the hardest problem we face in particle physics. It deserves
much more effort. I think the difficulty of this problem and the lack of a
``standard model'' of flavor are what have led to ETC's being in such
disfavor. Experiments will be of great help, possibly inspiring the right new
ideas. Certainly, experiments that will be done in this decade will rule out,
or vindicate, the ideas outlined in these lectures. That is an exciting
prospect!

\item{} More tractable is the problem of constructing a dynamical theory of
the top--quark mass that is natural, i.e., requires no fine--tuning of
parameters, and has no nearby Landau pole. Like topcolor--assisted
technicolor and top--seesaw models, such a theory is bound to have testable
consequences below 2--3~TeV. So hurry---before the experiments get done!

\item{} Neutrino masses are at least as difficult a problem as the top
mass. In particular, it is a great puzzle how ETC interactions could produce
$m_\nu \simle 10^{-7} m_e$. It seems unnatural to have to assume an extra
large ETC mass scale just for the neutrinos. Practically no thought has been
has been given to this problem. Is there some simple way to tinker with the
basic ETC mass--generating mechanism, some way to implement a seesaw
mechanism, or must the whole ETC idea be scrapped? The area is wide open.

\item{} My favorite problem is ``vacuum alignment'' and CP
violation~\cite{vacalign}. The basic idea is this: Spontaneous chiral
symmetry breaking implies the existence of infinitely many degenerate ground
states. These are manifested by the presence of massless Goldstone bosons
(technipions). The ``correct'' ground state, i.e., the one on which
consistent chiral perturbation theory for the technipions is to be carried
out, is the one which minimizes the vacuum expectation value of the explicit
chiral symmetry breaking Hamiltonian $\CH'$ generated by ETC. As Dashen
discovered, it is possible that an $\CH'$ that appears to conserve CP
actually violates it in the correct ground state. This provides a beautiful
dynamical mechanism for the CP violation we observe. Or it could lead to
disaster---strong CP violation, with a neutron electric dipole moment ten
orders of magnitude larger than its upper limit. This field of research is
just beginning in earnest. If the strong--CP problem can be controlled (there
is reason to hope that it can be!), there are bound to be new sources of CP
violation that are accessible to experiment.

\end{enumerate}

\section*{Acknowledgements}

I am grateful to Giulia Pancheri for inviting me to lecture at the 2000
Frascati Spring School. I owe much of my pleasure to the School's
enthusiastic students and their openness to the subversive science I told
them about. I give great thanks to the Frascati Spring School secretariat,
M.~Legramante and A.~Mantella, for many large and small assistances.  I thank
Yogi Srivastava for his invitation to visit the beautiful town and University
of Perugia and the opportunity to speak there. Lia's and Yogi's generous
support and hospitality leave me in their debt. I thank my colleagues at BU,
members of the Fermilab Run~II Strong Dynamics Workshop, and others for their
help and constructive comments. In particular, I thank Georges Azuelos,
Sekhar Chivukula, Estia Eichten, Andre Kounine, Greg Landsberg, Richard Haas,
Takanobu Handa, Robert Harris, Kaori Maeshima, Meenakshi Narain, Steve
Mrenna, Stephen Parke, Tongu\c c Rador, Francois Richard, Elizabeth Simmons,
and John Womersley. This research was supported in part by the
U.~S. Department of Energy under Grant~No.~DE--FG02--91ER40676.

\vfil\eject

\begin{thebibliography}{99}
%
%
\bibitem{smtheory} S.~L.~Glashow, {\it Nucl.~Phys.~}~{\bf 22}, 579 (1961);\\
S.~Weinberg, {\it Phys.~Rev.~Lett.}~{\bf 19}, 1264 (1967);\\
A.~Salam, in Proceedings of the 8th Nobel Symposium on Elementary Particle
Theory, Relativistic Groups and Analyticity, edited by N.~Svartholm, p.~367
(Almquist and Wiksells, Stockholm, 1968);\\
H.~Fritzsch, M.~Gell--Mann, and H.~Leutwyler, {\it Phys.~Lett.}~{\bf B47},
365 (1973);\\
D.~Gross and F.~Wilczek, {\it Phys.~Rev.~Lett.}~{\bf 30}, 1343 (1973);\\
H.~D.~Politzer, {\it Phys.~Rev.~Lett.}~{\bf 30}, 1346 (1973).
%
%
\bibitem{smexpt} R.~Cahn and G.~Goldhaber, {\it The Experimental
Foundations of Particle Physics}, (Cambridge University Press, 1989)
%
%
\bibitem{higgs} P.~W.~Anderson, {\it Phys.~Rev.}~{\bf 110},827 (1958); {\it
ibid.}, {\bf 130}, 439 (1963);\\
Y.~Nambu, {\it Phys.~Rev.}~{\bf 117}, 648 (1959);\\
J.~Schwinger, {\it Phys.~Rev.}~{\bf 125}, 397 (1962);\\
P.~Higgs, {\it Phys.~Rev.~Lett.}~{\bf 12}, 132 (1964);\\
F.~Englert and R.~Brout, {\it Phys.~Rev.~Lett.}~{\bf 13}, 321 (1964);\\
G.~S.~Guralnik, C.~R.~Hagen, and T.~W.~B.~Kibble, {\it Phys.~Rev.~Lett.}~{\bf
13}, 585 (1964).
%
%
\bibitem{pdg} Particle Data Group,
  http://pdg.lbl.gov/1999/contents\_sports.html.
%
%
\bibitem{quigg} For a recent review of the status of the standard model and
  concerns with its foundations, see C.~Quigg {\it The State of the Standard
  Model}, Lecture at Conference on the Physics Potential and Development of
  Muon Colliders and Neutrino Factories, San Francisco, December 15-17, 1999,
  hep-ph/0001145.
%
%
\bibitem{natural} K.~G.~Wilson, unpublished; quoted in L.~Susskind,
{\it Phys.~Rev.}~{\bf D20}, 2619 (1979);\\
G.~'t~Hooft, in {\it Recent Developments in Gauge Theories}, edited by
G.~'t~Hooft, et al. (Plenum, New York, 1980).
%
%
\bibitem{trivial} See, for example, R. Dashen and H. Neuberger, {\it
Phys.~Rev.~Lett.}~{\bf 50}, 1897 (1983) 1897;\\
J.~Kuti, L.~Lin, and Y.~Shen, {\it Phys.~Rev.~Lett.}~{\bf 61},678 (1988);\\
A.~Hasenfratz, et al.~{\it Phys.~Lett.}~{\bf B199}, 531 (1987);\\
G.~Bhanot and K.~Bitar, {\it Phys.~Rev.~Lett.}~{\bf 61}, 798 (1988).
%
%
\bibitem{rscdk} R.~S.~Chivukula and D.~Kominis, {\it Phys.~Lett.}~{\bf B304},
  152 (1993).
%
%
\bibitem{rscne} R.~S.~Chivukula and N.~Evans, {\it Phys.~Lett.}~{\bf B464},
  244 (1999); hep-ph/9907414.
%
%
\bibitem{tc} S.~Weinberg, Phys.~Rev.~{\bf D19}, 1277 (1979);\\
L.~Susskind, Phys.~Rev.~{\bf D20}, 2619 (1979).
%
%
\bibitem{kltasi} K~Lane, {\it An Introduction to Technicolor}, Lectures given
June~30--July~2 1993 at the Theoretical Advanced Studies Institute,
University of Colorado, Boulder, published in ``The Building Blocks of
Creation'', edited by S.~Raby and T.~Walker, p.~381, World Scientific (1994);
hep-ph/9401324.
%
%
\bibitem{rscreview}R.~S.~Chivukula, {\it Models of Electroweak Symmetry},
NATO Advanced Study Institute on Quantum Field Theory Perspective and
Prospective, Les Houches, France, 16--26 June 1998, hep-ph/9803219.
%
%
\bibitem{etceekl} E.~Eichten and K.~Lane, Phys.~Lett.~{\bf B90}, 125 (1980).
%
%
\bibitem{etcsd} S.~Dimopoulos and L.~Susskind, {\it Nucl.~Phys.~}~{\bf B155}
(1979) 237.
%
%
\bibitem{jjcn} K.~Johnson and R.~Jackiw, Phys.~Rev.~{\bf D8}, 2386 (1973);\\
  J.~Cornwall and R.~Norton, Phys.~Rev.~{\bf D8}, 3338 (1973);\\ also see
  M.~Weinstein, Phys.~Rev.~{\bf D8}, 2511 (1973), for an early discussion of
  dynamical electroweak symmetry breaking.
%
%
\bibitem{CPreview} See, e.g., R.~D.~Peccei, {\it QCD, Strong CP and Axions},
  hep-ph/9606475.
%
%
\bibitem{kdlhdp} K.~Lane, {\it Phys.~Rev.}~{\bf D10}, 2605 (1974);\\
H.~D.~Politzer {\it Nucl.~Phys.}~{\bf B117}, 397 (1976).
%
%
\bibitem{agchmg} A.~G.~Cohen and H.~M.~Georgi, Nucl.~Phys.~{\bf B314}, 7
(1989).
%
%
\bibitem{ehlq} E.~Eichten, I.~Hinchliffe, K.~Lane, and C.~Quigg,
  Rev.~Mod.~Phys~{\bf 56}, 579 (1984); Phys.~Rev.~{\bf D34}, 1547 (1986).
%
%
\bibitem{wtc}B.~Holdom, Phys.~Rev.~{\bf D24}, 1441 (1981);
Phys.~Lett.~{\bf B150}, 301 (1985);\\
T.~Appelquist, D.~Karabali, and L.~C.~R. Wijewardhana,
Phys.~Rev.~Lett.~{\bf 57}, 957 (1986);\\
T.~Appelquist and L.~C.~R.~Wijewardhana, Phys.~Rev.~{\bf D36}, 568
(1987);\\
K.~Yamawaki, M.~Bando, and K.~Matumoto, Phys.~Rev.~Lett.~{\bf 56}, 1335
(1986);\\
T.~Akiba and T.~Yanagida, Phys.~Lett.~{\bf B169}, 432 (1986).
%
%
\bibitem{hemc} K.~Lane, {\it Technicolor Signatures at the High Energy Muon
Collider}, Talk delivered at the workshop ``Studies on Colliders and Collider
Physics at the Highest Energies: Muon Colliders at 10 TeV to 100 TeV'',
Montauk, Long Island, NY, 27~September--1~October 1999; hep-ph/9912526.
%
%
\bibitem{ellisfcnc}J.~Ellis, M.~Gaillard, D.~Nanopoulos, and P.~Sikivie,
Nucl.~Phys.~{\bf B182}, 529 (1981).
%
%
\bibitem{pettests} B.~W.~Lynn, M.~E.~Peskin, and R.~G.~Stuart, in Trieste
Electroweak 1985, 213 (1985);\\
A.~Longhitano, Phys.~Rev.~{\bf D22}, 1166 (1980);
Nucl.~Phys.~{\bf B188}, 118 (1981);\\
R.~Renken and M.~Peskin, Nucl.~Phys.~{\bf B211}, 93 (1983);\\
M.~Golden and L.~Randall, Nucl.~Phys.~{\bf B361}, 3 (1990);\\
M.~E.~Peskin and T.~Takeuchi, Phys.~Rev.~Lett. {\bf 65}, 964 (1990);\\
B.~Holdom and J.~Terning, Phys.~Lett.~{\bf B247}, 88 (1990);\\
A.~Dobado, D.~Espriu, and M~J.~Herrero, Phys.~Lett.~{\bf B255}, 405
(1990);\\
H.~Georgi, Nucl.~Phys.~{\bf B363}, 301 (1991).
%
%
\bibitem{etcgim} S.~Dimopoulos, H.~Georgi, and S.~Raby, Phys.~Lett.~{\bf
B127}, 101 (1983);\\
S.--C.~Chao and K.~Lane, Phys.~Lett.~{\bf B159}, 135 (1985);\\
R.~S.~Chivukula and H.~Georgi, Phys.~Lett.~{\bf B188}, 99 (1987);\\
L.~Randall, Nucl.~Phys.~{\bf B403},122 (1993).
%
%
\bibitem{tctests}R.~S.~Chivukula, E.~H.~Simmons, and J.~Terning,
  Phys.~Lett.~{\bf B331}, 383 (1994);\\
T.~Takeuchi, W.~Loinaz, and A.~K.~Grant, Talk given at 13th Topical Conference
  on Hadron Collider Physics, Mumbai, India, 14-20 Jan 1999, hep-ph/9904207.
%
%
\bibitem{setc} T.~Appelquist, M.~B.~Einhorn, T.~Takeuchi, and 
L.~C.~R.~Wijewardhana, Phys.~Lett.~{\bf B220}, 223 (1989);\\
T.~Takeuchi, Phys.~Rev.~{\bf D40}, 2697 (1989);\\
V.A.~Miransky and K.~Yamawaki, Mod.~Phys.~Lett.~{\bf A4}, 129 (1989);\\
K.~Matumoto, Prog.~Theor.~Phys.~{\bf 81}, 277 (1989) ;\\
R.~S.~Chivukula, A.~G.~Cohen, and K.~Lane, Nucl.~Phys.~{\bf B343}, 554
(1990).
%
%
\bibitem{tombowick} T. Appelquist, M.~J.~Bowick, E.~Cohler, and A.~I.~Hauser,
  Phys.~Rev.~Lett.~{\bf 53}, 1523 (1984); Phys.~Rev.~{\bf D31}, 1676 (1985).
%
%
\bibitem{zbbth}R.~S.~Chivukula, S.~B.~Selipsky, and E.~H.~Simmons,
Phys.~Rev.~Lett.~{\bf 69}, 575 (1992);\\
R.~S.~Chivukula, E.~H.~Simmons, and J.~Terning,
Phys.~Lett.~{\bf B331}, 383 (1994), and references therein.
%
%
\bibitem{tctwohill}C.~T.~Hill, Phys.~Lett.~{\bf B345}, 483 (1995).
%
%
\bibitem{topcref}
C.~T. Hill, Phys.~Lett.~{\bf B266}, 419 (1991);\\
S.~P.~Martin, Phys.~Rev.~{\bf D45}, 4283 (1992);\\
{\it ibid}~{\bf D46}, 2197 (1992); Nucl.~Phys.~{\bf B398}, 359 (1993);\\
M.~Lindner and D.~Ross, Nucl.~Phys.~{\bf  B370}, 30 (1992);\\
R.~B\"{o}nisch, Phys.~Lett.~{\bf B268}, 394 (1991);\\
C.~T.~Hill, D.~Kennedy, T.~Onogi, H.~L.~Yu, Phys.~Rev.~{\bf D47}, 2940 
(1993).
%
%
\bibitem{alm} T.~Appelquist, K.~Lane, and U.~Mahanta, Phys.~Rev.~Lett.~{\bf
61}, 1553 (1988).
%
%
\bibitem{klglasgow} K.~Lane, Proceedings of the 27th International Conference
on High Energy Physics, edited by P.~J.~Bussey and I.~G.~Knowles, Vol.~II,
p.~543, Glasgow, June 20--27, 1994, hep-ph/9409304.
%
%
\bibitem{edrta}M.~Knecht and E.~de Rafael, Phys.~Lett.~{\bf B424}, 335 (1998);
hep-ph/9712457;\\
T.~Appelquist and F.~Sannino, Phys.~Rev.~{\bf D59}, 067702,
1999, hep-ph/9806409;\\ 
T.~Appelquist, P.~S.~Rodrigues, and F.~Sannino, Phys.~Rev.~{\bf D60}, 116007
(1999), hep-ph/9906555.
%
%
\bibitem{sfsr}S.~Weinberg, Phys.~Rev.~Lett.~{\bf 18}, 507 (1967);\\
K.~G.~Wilson, {\it Phys.~Rev.}~{\bf 179}, 1499 (1969);\\
C.~Bernard, A.~Duncan, J.~Lo~Secco, and S.~Weinberg,
Phys.~Rev.~{\bf D12}, 792 (1975).
%
%
\bibitem{topcondref}Y.~Nambu, in {\it New Theories in Physics}, Proceedings of
the XI International Symposium on Elementary Particle Physics, Kazimierz,
Poland, 1988, edited by Z.~Adjuk, S.~Pokorski and A.~Trautmann (World
Scientific, Singapore, 1989),Enrico Fermi Institute Report EFI~89-08
(unpublished);\\
V.~A.~Miransky, M.~Tanabashi, and K.~Yamawaki, Phys.~Lett.~{\bf
B221}, 177 (1989); Mod.~Phys.~Lett.~{\bf A4}, 1043 (1989);\\
W.~A.~Bardeen, C.~T.~Hill, and M.~Lindner, Phys.~Rev.~{\bf D41},
1647 (1990);\\
C.~T. Hill, Phys.~Lett.~{\bf B266}, 419 (1991);\\
S.~P.~Martin, Phys.~Rev.~{\bf D45}, 4283 (1992);\\
{\it ibid}~{\bf D46}, 2197 (1992); Nucl.~Phys.~{\bf B398}, 359 (1993);\\
M.~Lindner and D.~Ross, Nucl.~Phys.~{\bf  B370}, 30 (1992);\\
R.~B\"{o}nisch, Phys.~Lett.~{\bf B268}, 394 (1991);\\
C.~T.~Hill, D.~Kennedy, T.~Onogi, H.~L.~Yu, Phys.~Rev.~{\bf D47}, 2940 
(1993).
%
%
\bibitem{cdt} R.~S.~Chivukula, B.~Dobrescu, and J.~Terning, Phys.~Lett.~{\bf
B353}, 289 (1995), hep-ph/9503203.
%
%
\bibitem{tctwoklee}K.~Lane and E.~Eichten, Phys.~Lett.~{\bf B352}, 382
(1995), hep-ph/9503433.
%
%
\bibitem{tctwokl} K.~Lane, Phys.~Rev.~{\bf D54}, 2204 (1996),
  hep-ph/9602221;\\
K.~Lane, Phys.~Lett.~{\bf B433}, 96 (1998), hep-ph/9805254.
%
%
\bibitem{rscjt} R.~S.~Chivukula and J.~Terning, Phys.~Lett.~{\bf B385}, 209
  (1996), hep-ph/9606233.
%
%
\bibitem{ccs}R.~S.~Chivukula, A.~G.~Cohen and E.~H.~Simmons, Phys.~Lett. {\bf
B380}, 92 (1996), hep-ph/9603311;\\
M.~Popovic and E.~H.~Simmons, Phys.~Rev.~{\bf D58}, 095007
(1998), hep-ph/9806287.
%
%
\bibitem{multiklee} K.~Lane and E.~Eichten, Phys. Lett. {\bf B222}, 274
  (1989).
%
%
\bibitem{rsctriv} R.~S.~Chivukula and H.~Georgi, Phys.~Rev.~{\bf D58}, 115009
(1998), hep-ph/9806289.
%
%
\bibitem{seesaw}
B.~A.~Dobrescu and C.~T.~Hill,
Phs.~Rev.~Lett.~{\bf 81}, 2634 (1998), hep-ph/9712319;\\
R.~S.~Chivukula, B.~A.~Dobrescu, H.~M.~Georgi, and C.~T.~Hill,
Phys.~Rev.~{\bf D59}, 075003 (1999), hep-ph/9809470;\\
G.~Burdman and N.~Evans,
Phys.~Rev.~{\bf D59}, 115005 (1999), hep-ph/9811357.
%
%
\bibitem{jungle} H.~M.~Georgi and A.~K.~~Grant, {\it A Topcolor Jungle Gym},
hep-ph/0006050.
%
%
\bibitem{elw} E.~Eichten and K.~Lane, Phys.~Lett.~{\bf B388}, 803 (1996);
  hep-ph/9607213;\\ 
E.~Eichten, K.~Lane, and J.~Womersley, Phys.~Lett.~{\bf B405}, 305 (1997);
  hep-ph/9704455;\\ 
E.~Eichten, K.~Lane, and J.~Womersley, Phys.~Rev.~Lett.~{\bf 80}, 5489 (1998);
  hep-ph/9802368.
%
%
\bibitem{tcsm_singlet} K.~Lane, Phys.~Rev.~{\bf D60}, 075007 (1999),
  hep-ph/9903369. I refer to this paper as TCSM--1.
%
%
\bibitem{tcsm_octet} K.~Lane and S.~Mrenna, {\it Color--SU(3) Nonsinglet
    Technihadron Rates in the Technicolor Straw Man Model}, in preparation. I
    refer to this paper as TCSM--2.
%
%
\bibitem{pythia} T.~Sj\"ostrand, Comp.~Phys.~Com.~{\bf 82}, 74 (1994).
%
%
\bibitem{multiklrm} K.~Lane and M.~V.~Ramana, Phys.~Rev.~{\bf D44}, 2678
  (1991).
%
%
\bibitem{L3} The L3 Collaboration, {\it Search for Technicolor Production at
    LEP}, L3 Note~2428, submitted to the International Europhysics Conference
    High Energy Physics~99, Tampere, Finland 15--21 July 1999;
    http://l3www.cern.ch/conferences/EPS99.
%
%
\bibitem{delphi} The DELPHI Collaboration, DELPHI Note 2000-088 CONF~387,
  paper No.~375, presented at the XXXth International Conference on High
  Energy Physics, July 2000, Osaka, Japan.
%
\bibitem{cdfwpi} The CDF Collaboration, Phys.~Rev.~Lett.~{\bf 84}, 1110 (2000).
%
%
\bibitem{cdfgpi} The CDF Collaboration, Phys.~Rev.~Lett.~{\bf 83}, 3124
(1999).
%
%
\bibitem{handa} I am grateful to T.~Handa of CDF for supplying these plots..
%
%
\bibitem{dzeroee} D.~Toback {\it New Phenomena~II: Recent Results from the
    Fermilab Tevatron} (for the CDF and D\O\ collaborations), Proceedings of
    the 35th Rencontres de Moriond: Electroweak Interactions and Unified
    Theories (Moriond 2000), hep-ex/0005020.
%
%
\bibitem{meena} I am grateful to M.~Narain of D\O\ for this plot.
%
%
\bibitem{atlastdr} Atlas Physics Technical Design Report, Chapter~21
http://atlasinfo.cern.ch/ Atlas/GROUPS/PHYSICS/TDR/access.html, (1999).
%
%
\bibitem{fs} E.~Farhi and L.~Susskind, Phys.~Rev.~{\bf D20}, 3404 (1979).
%
%
\bibitem{cdfed} The CDF Collaboration, Phys.~Rev.~Lett.~{\bf 82}, 3206
(1999).
%
%
\bibitem{cdfnd} The CDF Collaboration, (T.~Affolder et al.), {\it Search for
Second and Third Generation Leptoquarks Including Production via Technicolor
Interactions in $p\ol p$ Collisions at $\ecm = 1.8\,\tev$},
FERMILAB-PUB-00-073-E, Apr 2000, submitted to Physical Review Letters,
hep-ex/0004003.
%
%
\bibitem{cdfjets} The CDF Collaboration, Phys.~Rev.~{\bf D55}, R5263
(1997).
%
%
\bibitem{cdfbbjets} The CDF Collaboration, Phys.~Rev.~Lett.~{\bf 82}, 2038
    (1999).
%
%
\bibitem{vacalign} R.~F.~Dashen, Phys.~Rev.~{\bf D3}, 1879 (1971);\\
E.~Eichten, K.~Lane, and J.~Preskill, Phys.~Rev.~Lett.~{\bf 45},
  225 (1980);\\
K.~Lane, Physica Scripta {\bf 23}, 1005 (1981);\\
J.~Preskill, Nucl.~Phys.~{\bf B177}, 21 (1981); \\
M.~E.~Peskin, Nucl.~Phys.~{\bf B175}, 197 (1980);\\
K.~Lane, T.~Rador, and E.~Eichten, Phys.~Rev.~{\bf D62}, 015005 (2000);
hep-ph/0001056.
%
%
\bibitem{ttlimit} The D\O\ Collaboration, Phys.~Rev.~Lett.~{\bf 79}, 1203
  (1997); hep-ex/9704015;\\
The D\O\ Collaboration, Phys.~Rev.~{\bf D60}, 012001 (1997);\\
F.~Ptohos (for the CDF collaboration), proceedings of the International
Europhysics Conference on High Energy Physics 99, Tampere, Finland, July 17,
1999.
%
%
\end{thebibliography}
\end{document}